\documentclass[preprint]{elsarticle}

\usepackage{amssymb,amsmath,amsthm,graphicx,amscd}
\usepackage[mathscr]{eucal}
\usepackage{enumerate,color,verbatim,multirow,comment}
\usepackage{bm}

\journal{Annals of Physics}

\begin{document}

\baselineskip=18pt
\numberwithin{equation}{section}

\allowdisplaybreaks

\begin{frontmatter}

\title{Entanglement Structure of an Open System \\ of $N$ Quantum Oscillators: \\ II. Strong Disparate Couplings $N=3$}

\author{J.-T. Hsiang}
\ead{cosmology@gmail.com}
\address{Department of Physics, National Dong Hwa University, Hualian, Taiwan}
\author{Rong Zhou}
\ead{zhour@umd.edu}
\address{Joint Quantum Institute and Maryland Center for Fundamental Physics, \\ University of Maryland, College Park, Maryland 20742}
\author{B.~L.~Hu}
\ead{blhu@umd.edu}
\address{Joint Quantum Institute and Maryland Center for Fundamental Physics, \\ University of Maryland, College Park, Maryland 20742}

%\date{Readable version by Thursday June 6, submit to arXiv June 14, 2013}

%%%%%%%%%%%%%%%%%%%%%%%%%%%%%%%%%%%%%

\begin{abstract}
In this paper we study a system of $N$ coupled quantum oscillators interacting with
each other directly with varying coupling strengths and indirectly through linear couplings to a scalar massless quantum field as its environment. The influence of the quantum field on the system is calculated with the use of the influence functional formalism.  We take the direct route of seeking solutions to the evolutionary operator of the reduced density matrix for the derivation of the correlation functions. They are then used to construct the covariance matrix which we use to perform an analysis of the structure of quantum entanglement in the open system at a stationary state. To see the physical features more explicitly we specialize to a system of three quantum coupled oscillators placed at the vertices of a equilateral triangle and allowed to have  disparate pairwise couplings. We analyze the entanglement between one oscillator and the other two with equal (symmetric) and unequal (asymmetric) coupling strengths. As an illustration we use the results for these two different configurations to address two representative issues in macroscopic quantum  phenomena.  We also mention possible extensions of our work and applications of our analysis and results to issues in some current areas of research in quantum thermodynamics and mesoscopic quantum systems.
\end{abstract}


\end{frontmatter}

%\maketitle

%\newpage

\tableofcontents

\section{Introduction}

In our first paper on this subject matter \cite{NOS1} we  studied a model of $N$ quantum oscillators (QO) located at different fixed positions in space, which do not interact with each other directly but only through  weak couplings to a common scalar massless quantum field. After coarse-graining the quantum field we derived the late-time covariance matrix of the open system of the $N$ oscillators with field-induced coupling and analyzed the structure of their quantum entanglement behavior for the case of $N=3$. We show that in the weak coupling limit the correlations adopt a simple pairwise structure, in that the correlation between any two QO is ignorant of the presence of the other QO. The effects of field-induced interaction on quantum entanglement between two harmonic oscillators have been studied earlier in \cite{LinHu09}. One interesting new  feature shown there is the distance dependence of quantum entanglement, since the coupling is mediated by a quantum field whose influence varies with time and space.  The way how one oscillator's presence alters the entanglement of the other two  was used  in \cite{NOS1} to capture a novel notion called `entanglement domain'.  In this paper we study the entanglement structure of the $N$ quantum oscillator system but in addition to each oscillator's weak coupling with a common quantum scalar field we allow also direct pairwise couplings with varying strength. The salient features based on detailed studies of the $N=3$ case turn out to be quite different from the case of $N$ oscillators with only field-induced coupling.  The first entanglement feature, namely, that the correlation between any two QO are ignorant of the presence of the other QO, is no longer true. The second feature, namely, distance dependence of entanglement due to each oscillator's coupling to a quantum field, is now no longer prominent, being overwhelmed by the direct couplings. We see that each case  studied in these two papers captures one  aspect of the overall behavior of the system of NOS (with $N=3$ studied in detail here): While the first case may depict neutral atoms the latter case depicts ions (in an open quantum field rather than in a cavity or in a trap). The idealized model-study here, though not a replica of actual experimental setups, can nonetheless provide us with a clear theoretical understanding and quantitative description of the entanglement structure of such commonly encountered continuous-variable quantum systems.

Quantifying the entanglement structure of a many-body quantum system can facilitate investigations into related issues in several current research areas.  For example, the results reported in this paper for a system of $N$ coupled oscillators with direct yet disparate couplings  can help  shed light on issues in \textit{Macroscopic quantum phenomena} (MQP) (See, e.g., \cite{MQP1,MQP2,MQP3,ChenMQM} and references therein).
Entanglement being an exclusively quantum feature \cite{Schrodinger}, it is natural to ask how to use entanglement as a signifier or marker of the existence of, or further, a quantitative measure of, the quantum features and behaviors in a macroscopic quantum system.  On this issue it is important to recognize that (see, e.g., \cite{E/QG}) there exist many \textit{levels of structure }in a composite body and only by judicious choice of an appropriate set of \textit{collective variables} can one give a good description of the dynamics of a specific level of structure. These are often characterized by strong coupling between fine-grained elements or constituents in one particular level of structure (e.g. the level of quarks / gluons or the level of nucleons) and weaker coupling between the elements in a coarser-grained level of structure (e.g., atoms or molecules).  Capturing the quantum features of a macroscopic object  is greatly facilitated by the existence and functioning of these collective variables depicting a specific level of structure. Under what conditions can one capture the quantum dynamics of an $N$ oscillator system by its center of mass (CoM), the so-called ``CoM Axiom", was investigated earlier by Chou et al \cite{CHYu} who derived a sufficient condition for this to hold.

A recent essay  \cite{HuSubasi} (see also cited work therein e.g., \cite{Martins})  showed the marked difference between entanglement within one specific level of structure and entanglement between different levels of structure,  and identified which types would scale with $N$. These two aspects, namely, the role of the CoM in a many-body system and how the level-of-structure enter in the manifestation of macroscopic quantum phenomena, can be examined by using the entanglement structure results obtained here for a three-oscillator system with disparate coupling strengths.

The paper is organized as follows: In the next section we derive a set of Langevin equations for $N$ coupled quantum oscillators interacting with a quantum scalar field based on the influence functional formalism approach. This method first introduced by Feynman and Vernon \cite{FeyVer} for the study of quantum Brownian motion has been applied to the derivation of Markovian and non-Markovian master equations for oscillator baths \cite{CalLeg83,HPZ} and for a quantum field environment \cite{HM94,RHA96,JH02}. (For an alternative treatment of a $N$ oscillator system, see \cite{FlemingNHO} and references therein.)  Here we take the direct route of seeking solutions to the evolutionary operator of the reduced density matrix given formally by Grabert et al \cite{GSI88} for the derivation of correlation functions. (For an alternative derivation of the non-Markovian master equation and finding their solutions, see \cite{CRV03,FRH11} and references therein.) In Sec. 3 these are used to construct the covariance matrix of this system which is at the heart of our analysis of quantum entanglement. In Sec. 4 we specialize to $N=3$ with disparate coupling and analyze the symmetric and asymmetric configurations described above.  In Sec. 5 we summarize our results and briefly describe their implications for some key issues in MQP and applications in other areas. The Appendix contains more details about symmetric Gaussian systems.

%\newpage

\section{A system of $N$ interacting detectors}

We consider a system of $N$ interacting $(i = 1, .., N)$ detectors where each detector's internal degree of freedom $\chi^{(i)}(t)$ is  described by a one-dimensional harmonic oscillator. These detectors are allowed to have direct coupling via their internal degrees of freedom, and the coupling is assumed to be at most quadratic in the internal degrees of freedom. In addition, each oscillator is coupled to one and the same massless scalar field $\phi(\mathbf{x},t)$ which acts as an environment to the $N$ oscillator system. As such there will be indirect interactions between the oscillators mediated by the shared bath. The entanglement features of such a system with only field-induced interaction was studied in our first paper \cite{NOS1}.

For later purpose when we will consider a system of $N$ moving detectors (such as that studied before in \cite{RHA96}) we can allow the detectors to be in motion with the trajectory of the $i^{th}$ detector specified by $\mathbf{z}^{(i)}(t)$ \cite{HHL}. We assume that all the oscillators have the same mass $m$, the same bare frequency $\omega_{0}$, and are acted upon by a  harmonic potential of the same form $V(\chi^{(i)})=m\omega_{0}^{2}\chi^{(i)}{}^{2}/2$.

The Lagrangian for the afore-specified system is given by
\begin{align}
	S[\chi,\phi]&=\int^{t}\!ds\sum_{i=1}^{N}\Bigl[\frac{m}{2}\,\dot{\chi}^{(i)2}(s)-V(\chi^{(i)})-\sum_{j> i}^{N}V_{ij}(\chi)\Bigr]\notag\\
	&\qquad\qquad+\int^{t}\!ds\int\!d^{3}\mathbf{x}\sum_{i=1}^{N}j^{(i)}(\mathbf{x},s)\phi(\mathbf{x},s)\notag\\
	 &\qquad\qquad\qquad+\frac{1}{2}\int^{t}\!ds\int\!d^{3}\mathbf{x}\Bigl[\partial_{\mu}\phi(\mathbf{x},s)\Bigr]\Bigl[\partial^{\mu}\phi(\mathbf{x},s)\Bigr]\,,\label{E:ypowpea}
\end{align}
where $j^{(i)}(\mathbf{x},s)=g_{i}\,\chi^{(i)}(s)\,\delta^{(3)}(\mathbf{x}-\mathbf{z}^{(i)}(s)\,)$ describes the source that accounts for the interaction between the internal degree of freedom $\chi^{(i)}$ of the $i^{th}$ detector and the environmental scalar field $\phi$ with coupling strength $g_{i}$ which we will assume to be weak. The direct coupling between the detectors are given by the interaction potential $V_{ij}$ of the form
\begin{equation}
	V_{ij}(\chi)=\frac{m\sigma_{ij}}{2}\Bigl[\chi^{(i)}(s)-\chi^{(j)}(s)\Bigr]^{2}\,,\qquad\sigma_{ij}=\sigma_{ji}\,,
\end{equation}
with the coupling strength $\sigma_{ij}$ which is positive and can be strong. The partial derivative $\partial_{\mu}$ in the action of the free scalar field represents $\partial_{\mu}=(\partial_{t},\pmb{\nabla})$ and we choose the Minkowski metric $\eta_{\mu\nu}$ with the convention $\eta_{\mu\nu}=\operatorname{diag}(+,-,-,-)$.

Since we are primarily interested in the dynamics of the internal degrees of freedom $\chi^{(i)}$ of the detectors, obtaining the evolution of their reduced density matrix of $\varrho_{\chi}$ constructed by integrating over the scalar field environment will serve our purpose. This enables us to find the corresponding time evolution of the expectation value of an operator $\mathcal{O}$ associated with the detector's internal degrees of freedom by
\begin{equation}\label{E:rnkenas}
	\langle\mathcal{O}\rangle=\operatorname{Tr}_{\chi}\Bigl\{\varrho_{\chi}\mathcal{O}\Bigr\}\,,
\end{equation}
where $\operatorname{Tr}_{\chi}$ means taking the trace over a complete set of $\chi$ states. For simplicity we use $\chi$ to represent the collection of the variables $\{\chi^{(i)}\}$. The time evolution of the reduced density operator, on the other hand, is given by
\begin{equation}\label{E:woejs}
	\varrho_{\chi}(t)=\operatorname{Tr}_{\phi}\Bigl\{U(t,t_{a})\varrho(t_{a})U^{-1}(t,t_{a})\Bigr\}\,.
\end{equation}
Here $\varrho(t_{a})$ is the full density operator at the initial time $t_{a}$, which describes the initial states of the combined systems $\chi$ and $\phi$, and $U(t,t_{a})$ is the unitary time evolution operator for the complete system governed by the action \eqref{E:ypowpea}. The trace operator $\operatorname{Tr}_{\phi}$ acts on the degrees of freedom of the background scalar field. Thus the reduced density operator $\varrho_{\chi}$ alone has offered us sufficient information about the time evolution of the reduced system under the influence of the background scalar field.

Here we assume at the initial time $t_{a}$, the state of the total system can be factorized into a direct product of the initial state $\varrho_{\chi}$ of $\chi$ and the counterpart $\varrho_{\beta}$ of the scalar field $\phi$,
\begin{equation}\label{E:rnkjwa}
	\varrho(t_{a})=\varrho_{\chi}(t_{a})\otimes\varrho_{\beta}\,,
\end{equation}
each of which explicitly takes the form
\begin{align}
	 \varrho_{\chi}(\chi^{(i)}_{a},\chi'^{(i)}_{a};t_{a})&=\left(\frac{1}{\pi\varsigma^{2}}\right)^{1/2}\exp\left[-\frac{1}{2\varsigma^{2}}\bigl(\chi^{(i)2}_{a}+\chi'^{(i)2}_{a}\bigr)\right]\,,\label{E:erjdkjnds}\\
	\varrho_{\beta}(\phi_{a},\phi'_{a};t_{a})&=\langle\phi_{a}\vert e^{-\beta\,H_{\phi}[\phi]}\vert\phi'_{a}\rangle\,.
\end{align}
The parameter $\varsigma$ is the width of the wavepacket associated with $\chi^{(i)}$, and $H_{\phi}[\phi]$ is the free scalar field Hamiltonian. The variables $\chi_{a}$, $\chi_{b}$ represent $\chi(t_{a})$ and $\chi(t_{b})$ respectively. We will use this shorthand notation for other variables. The initial state \eqref{E:rnkjwa} describes the system  (with variables $\chi^{(i)}$) prepared as wavepackets of equal width $\varsigma$ and the environment as that of a thermal scalar field at temperature $\beta^{-1}$. At $t_{a}$ the NOS is brought in contact with the field environment. The evolution of the combined system + environment will be governed by the unitary evolution operator $U(t_{b},t_{a})$ till the final time $t_{b}$ according to \eqref{E:woejs}. Although the initial state \eqref{E:rnkjwa} is na\"ively simplified and a more sophisticated and physically realistic  state has been proposed by~\cite{GSI88, FRH11, RP97}, it is amply sufficient to describe to the late-time dynamics of the reduced system. The possible artifact induced by a factorized initial state is significant only within a very short time from $t_a$ determined by the highest frequency cutoff in the combined system~\cite{HPZ}.

With these provisions the final reduced density matrix for the internal degrees of the detectors can be calculated exactly with the help of the influence functional formalism~\cite{FeyVer, GSI88}. Once we express \eqref{E:woejs} in terms of path integrals, we have
\begin{align}
	&\quad\varrho_{\chi}\bigl(\chi_{b},\chi'_{b};t_{b}\bigr)\notag\\
	 &=\int_{-\infty}^{\infty}\!d\chi^{\vphantom{'}}_{a}\!\int_{-\infty}^{\infty}\!d\chi'_{a}\!\int_{\chi^{\vphantom{'}}_{a}}^{\chi_{b}}\mathcal{D}\chi_{+}\!\int_{\chi'_{a}}^{\chi'_{b}}\mathcal{D}\chi_{-}\!\int_{-\infty}^{\infty}\!d\phi_{b}\int_{\phi_{a}}^{\phi_{b}}\mathcal{D}\phi_{+}\!\int_{\phi'_{a}}^{\phi_{b}}\mathcal{D}\phi_{-}\!\int_{-\infty}^{\infty}\!d\phi_{a}\!\int_{-\infty}^{\infty}\!d\phi'_{a}\;\notag\\
	 &\qquad\times\varrho\bigl(\chi^{\vphantom{'}}_{a},\phi_{a};\chi'_{a},\phi'_{a};t_{a}\bigr)\exp\biggl\{i\,S\bigl[\chi_{+},\phi_{+}\bigr]-i\,S\bigl[\chi_{-},\phi_{-}\bigr]\biggr\}\,,\label{E:wondkwqooq}
\end{align}
which has a Gaussian integrand. After carrying out the multiple integrals, we arrive at
\begin{equation}\label{E:lmlwe}
	 \varrho_{\chi}\bigl(r_{b},q_{b};t_{b}\bigr)=\int_{-\infty}^{\infty}\!dr_{a}dq_{a}\;J\bigl(r_{b},q_{b},t_{b};r_{a},q_{a},t_{a}\bigr)\,\varrho_{\chi}\bigl(r_{a},q_{a};t_{a}\bigr)\,,
\end{equation}
with $q(t)=\chi_{+}(t)-\chi_{-}(t)$ and $r(t)=\bigl[\chi_{+}(t)+\chi_{-}(t)\bigr]/2$. The evolutionary operator $J$ essentially describes how the reduced density operator $\varrho_{\chi}$ of the detectors evolves with time. It is given by
\begin{align}\label{E:rijwnewk}
	 J\bigl(r_{b},q_{b},t_{b};r_{a},q_{a},t_{a}\bigr)&=N(t_{b},t_{a})\,\exp\Bigl\{i\,\overline{S}_{CG}\bigl[\chi_{+},\chi_{-}\bigr]\Bigr\}\,,
\end{align}
where the coarse-grained effective action $S_{CG}$ is defined by
\begin{equation}\label{E:ernksjds}
	S_{CG}[\chi_{+},\chi_{-}]=S_{\chi}\bigl[\chi_{+}\bigr]-S_{\chi}\bigl[\chi_{-}\bigr]-i\,\ln {\cal F}\bigl[\chi_{+},\chi_{-}\bigr]\,,
\end{equation}
with the action of $\chi$ given by
\begin{equation}
	S_{\chi}[\chi]=\int^{t}\!ds\sum_{i=1}^{N}\Bigl[\frac{m}{2}\,\dot{\chi}^{(i)2}(s)-V(\chi^{(i)})-\sum_{j> i}^{N}V_{ij}(\chi)\Bigr]\,,
\end{equation}
and the influence functional ${\cal F}$ given by
\begin{align}\label{E:weijwsja}
	&\quad {\cal F}[\chi_{+},\chi_{-}]\\
	 &=\exp\left[i\int_{t_{a}}^{t_{b}}\!ds\!\int_{t_{a}}^{t_{b}}ds'\sum_{i,\,j}g_{i}g_{j}\left\{q^{(i)}(s)G^{ij}_{R}(s,s')r^{(j)}(s')+\frac{i}{2}\,q^{(i)}(s)G^{ij}_{H}(s,s')q^{(j)}(s')\right\}\right]\,.\notag
\end{align}
We use an overbar above the action or the influence functional to connote that they are evaluated along the classical trajectory of $\chi_{+}$, $\chi_{-}$. The normalization constant $N(t_{b},t_{a})$ in \eqref{E:rijwnewk} will be determined by the completeness relation.

\subsection{Influence functional}

The influence functional ${\cal F}$ merits some discussion. It essentially summarize the effects of the environment scalar field on the internal degrees of freedom $\chi$ after we trace out the scalar field in \eqref{E:wondkwqooq}. It contains two kernel functions $G_{R}(s,s')$ and $G_{H}(s,s')$, which are respectively the dissipation kernel and the fluctuation kernel~\cite{JH02, HWL08},
\begin{align}
	 G^{ij}_{R}(s,s')&=i\,\theta(s-s')\,\operatorname{Tr}_{\phi}\Bigl(\Bigl[\phi(\mathbf{z}^{(i)}(s),s),\phi(\mathbf{z}^{(j)}(s'),s')\Bigr]\times\varrho_{\beta}\Bigr)\,,\\
	 G^{ij}_{H}(s,s')&=\frac{1}{2}\,\operatorname{Tr}_{\phi}\Bigl(\Bigl\{\phi(\mathbf{z}^{(i)}(s),s),\phi(\mathbf{z}^{(j)}(s'),s')\Bigr\}\times\varrho_{\beta}\Bigr)\,,\label{E:enrkea}
\end{align}
where $\operatorname{Tr}_{\phi}$ is the trace taken over the field variables. While the noise kernel represents the coarse-grained effect of the environment as stochastic source(s) and the dissipation kernel captures the backreaction effect of the coarse-grained environment on the detectors' dynamics. Ostensibly not to be arbitrarily assigned by hand they are bound by a set of fluctuation-dissipation relations which should be well-known and in the  general case also a set of propagation-correlation relation which are lesser known~\cite{RHA96}. Furthermore, they mediate the response of one detector to the others both by direct couplings amongst the detector and even in the case when there is no direct coupling since the trajectory of each detector affects the field which in turn affects the other detectors. Therefore for such a $N$-detector system, the dynamics of a NOS is highly non-Markovian. Note incidentally they should not be mistaken to be tensorial quantities, the superscripts $ij$ are merely labels for the action of the $j^{th}$ detector on the $i^{th}$ detector.

It proves more convenient later if we cast the evolutionary operator $J$ in \eqref{E:rijwnewk} in the matrix notation and write it explicitly in terms of the initial and the final values of $q^{(i)}$, $r^{(i)}$. Let us denote the column matrix $\pmb{\chi}$ by $\pmb{\chi}=(\chi^{(1)},\chi^{(2)},\ldots,\chi^{(N)})^{T}$ and the coupling matrix $\mathbf{g}$ as $\mathbf{g}=\operatorname{diag}(g_{1}\,g_{2}\,\cdots\,g_{N})$. The action $S_{\chi}$ then assumes a compact form
\begin{equation}
	 S_{\chi}\bigl[\chi\bigr]=\int^{t_{b}}_{t_{a}}\!ds\;\Bigl[\frac{m}{2}\,\dot{\pmb{\chi}}^{T}\cdot\dot{\pmb{\chi}}-\frac{m}{2}\,\pmb{\chi}^{T}\cdot\pmb{\Omega}^{2}\cdot\pmb{\chi}\Bigr]\,,
\end{equation}
where $T$ in the superscript means the transpose of a matrix. The $N\times N$ interaction matrix $\pmb{\Omega}^{2}$, given by
\begin{equation}\label{E:jeaiekjna}
	\pmb{\Omega}^{2}=\begin{pmatrix}
		\omega_{0}^{2}+\displaystyle\sum^{N}_{k=1}{}'\sigma_{1k} &-\sigma_{12} &-\sigma_{13} &\cdots &-\sigma_{1N}\\[8pt]
		-\sigma_{21} &\omega_{0}^{2}+\displaystyle\sum^{N}_{k=1}{}'\sigma_{2k} &-\sigma_{23} &\cdots &-\sigma_{2N}\\[8pt]
		-\sigma_{31} &-\sigma_{32} &\ddots &\cdots &-\sigma_{3N}\\[8pt]
		\vdots &\vdots &\vdots &\ddots &\vdots\\[8pt]
		-\sigma_{N1} &-\sigma_{N2} &-\sigma_{N3} &\cdots &\omega_{0}^{2}+\displaystyle\sum^{N}_{k=1}{}'\sigma_{Nk}
	\end{pmatrix}\,,
\end{equation}
essentially indicates the strength of direct coupling between the internal degrees of freedom of the different detectors. The prime next to the summation, for example,
\begin{equation*}
	\displaystyle\sum^{N}_{k=1}{}'\sigma_{jk}
\end{equation*}
means that the summation index $k$ runs from $1$ to $N$ except for $k=j$.

If we vary the effective action $S_{CG}$ \eqref{E:ernksjds} with respect to $\mathbf{q}$ or $\mathbf{r}$, it produces two sets of equations of motion for the time evolution of $\mathbf{r}$, $\mathbf{q}$, respectively. In particular, the equation of motion for the real part of $\mathbf{r}$ is of special importance
\begin{align}\label{E:howswqkz}
	 m\,\ddot{\mathbf{r}}_{R}(s)+m\,\pmb{\Omega}^{2}\cdot\mathbf{r}_{R}(s)-\int_{0}^{s}\!ds'\;\mathbf{G}_{R}(s,s')\cdot\mathbf{r}_{R}(s')&=0\,,
\end{align}
where for simplicity we have let the initial time $t_{a}=0$ and the final time $t_{b}=t$. This equation is highly coupled due to both off-diagonal elements of the interaction matrix $\pmb{\Omega}^{2}$ and the non-local integral expression in the equation of motion. In general even if  a suitable basis is found to diagonalize the interaction matrix, its non-Markovian nature still render this equation very much tangled. Later in Sec.~\ref{S:rnfekjna} we will focus on a special case when the equation of motion \eqref{E:howswqkz} can be totally decoupled.

In general the solution to this equation can formally be expressed in terms of the fundamental solutions $\mathbf{D}_{1}$, $\mathbf{D}_{2}$, which satisfy the same equation of motion \eqref{E:howswqkz} and the  conditions
\begin{align*}
	\mathbf{D}_{1}(0)&=\mathbf{I}_{N}\,,&\dot{\mathbf{D}}_{1}(0)&=0\,,\\
	\mathbf{D}_{2}(0)&=0\,,&\dot{\mathbf{D}}_{2}(0)&=\mathbf{I}_{N}\,.
\end{align*}
The matrix $\mathbf{I}_{N}$ is an $N\times N$ identity matrix. We immediately see from \eqref{E:howswqkz} that these fundamental solutions contain information of dissipation effects due to the backreaction of the environment. The fundamental solutions to $\mathbf{r}_{R}$ can also be used to construct the solutions of $\mathbf{q}$, so the general solutions of $\mathbf{r}_{R}$ and $\mathbf{q}$ are both given by
\begin{align}
	 \mathbf{r}_{R}(s)&=\Bigl[\mathbf{D}_{1}(s)-\mathbf{D}_{2}(s)\cdot\mathbf{D}^{-1}_{2}(t)\cdot\mathbf{D}_{1}(t)\Bigr]\cdot\mathbf{r}_{a}+\mathbf{D}_{2}(s)\cdot\mathbf{D}^{-1}_{2}(t)\cdot\mathbf{r}_{b}\,,\label{E:uiewjs}\\
	 \mathbf{q}(s)&=\Bigl[\mathbf{D}_{1}(t-s)-\mathbf{D}_{2}(t-s)\cdot\mathbf{D}^{-1}_{2}(t)\cdot\mathbf{D}_{1}(t)\Bigr]\cdot\mathbf{q}_{b}+\mathbf{D}_{2}(t-s)\cdot\mathbf{D}^{-1}_{2}(t)\cdot\mathbf{q}_{a}\,,
\end{align}
respectively. Hereafter we will suppress the subscript $R$ in $\mathbf{r}_{R}$ to simplify notations. Note that the velocities $\dot{\mathbf{r}}_{a}$, $\dot{\mathbf{r}}_{b}$ at the initial and the final moment can also be given in terms of the fundamental solutions and $\mathbf{r}_{a}$, $\mathbf{r}_{b}$.

With these results ready at hand, the evolutionary operator becomes
\begin{align}
	&\quad J(\mathbf{q}_{b},\mathbf{r}_{b},t_{b}=t;\mathbf{q}_{a},\mathbf{r}_{a},t_{a}=0)\\
	 &=\exp\left\{i\,m\,\mathbf{q}^{T}_{b}\cdot\dot{\mathbf{r}}_{b}-i\,m\,\mathbf{q}^{T}_{a}\cdot\dot{\mathbf{r}}_{a}-\frac{1}{2}\Bigl[\mathbf{q}_{a}^{T}\cdot\mathbf{A}\cdot\mathbf{q}_{a}+\mathbf{q}_{b}^{T}\cdot\mathbf{B}\cdot\mathbf{q}_{b}+\mathbf{q}_{a}^{T}\cdot\mathbf{C}\cdot\mathbf{q}_{b}+\mathbf{q}^{T}_{b}\cdot\mathbf{C}^{T}\cdot\mathbf{q}_{a}\Bigr]\right\}\,.\notag
\end{align}
The matrices $\mathbf{A}$, $\mathbf{B}$ and $\mathbf{C}$ are defined by
\begin{align}
	 \mathbf{A}(t)&=\int_{0}^{t}\!ds\!\int_{0}^{t}\!ds'\;\pmb{\mu}^{T}(s)\cdot\mathbf{g}\cdot\mathbf{G}_{H}(s,s')\cdot\mathbf{g}\cdot\pmb{\mu}(s')\,,\label{E:rehkwd}\\
	 \mathbf{B}(t)&=\int_{0}^{t}\!ds\!\int_{0}^{t}\!ds'\;\pmb{\nu}^{T}(s)\cdot\mathbf{g}\cdot\mathbf{G}_{H}(s,s')\cdot\mathbf{g}\cdot\pmb{\nu}(s')\,,\\
	 \mathbf{C}(t)&=\int_{0}^{t}\!ds\!\int_{0}^{t}\!ds'\;\pmb{\mu}^{T}(s)\cdot\mathbf{g}\cdot\mathbf{G}_{H}(s,s')\cdot\mathbf{g}\cdot\pmb{\nu}(s')\,.
\end{align}
They highlight the backreaction effects due to the fluctuations of the scalar field environment, in contrast with the information about the dissipation backreaction effects wrapped up in the fundamental solutions $\mathbf{D}_{i}$. The matrix $\mathbf{G}_{H}$ is composed of elements \eqref{E:enrkea}. The two matrix functions $\pmb{\mu}(s)$, $\pmb{\nu}(s)$ are constructed by
\begin{align*}
	\pmb{\mu}(s)&=\mathbf{D}_{2}(t-s)\cdot\mathbf{D}^{-1}_{2}(t)\,,\\
	\pmb{\nu}(s)&=\mathbf{D}_{1}(t-s)-\mathbf{D}_{2}(t-s)\cdot\mathbf{D}^{-1}_{2}(t)\cdot\mathbf{D}_{1}(t)\,,
\end{align*}
where $\mathbf{D}_{1,2}$ are the fundamental solutions introduced in \eqref{E:howswqkz}.

By \eqref{E:lmlwe}, and the explicit expression of the evolutionary operator $J$, we can obtain the reduced density matrix for $\chi$ at the final time, from which we are able to compute the expectation values of the operators associated with $\chi$ at that moment according to \eqref{E:rnkenas}. This enables us to investigate the non-equilibrium dynamics of the reduced system $\chi$ under the influence of the scalar field environment.

\subsection{Correlation functions}

For the stated purpose of this work the evolution of the correlations between components of $\pmb{\chi}_{b}$ and their conjugate momenta $\mathbf{p}_{b}$ are of special interest.  We summarize the results here, %~\cite{}
\begin{align}
	 \langle\,\pmb{\chi}_{b}^{(l)}\pmb{\chi}_{b}^{(m)}\rangle&=\int\!d\mathbf{r}_{b}d\mathbf{q}_{b}\int\!d\mathbf{r}_{a}d\mathbf{q}_{a}\;\delta^{(N)}(\mathbf{q}_{b})\Bigl[\mathbf{r}_{b}^{(l)}+\frac{1}{2}\,\mathbf{q}_{b}^{(l)}\Bigr]\Bigl[\mathbf{r}_{b}^{(m)}+\frac{1}{2}\,\mathbf{q}_{b}^{(m)}\Bigr]\notag\\
	&\qquad\qquad\qquad\qquad\qquad\qquad\qquad\qquad\times J(\mathbf{r}_{b},\mathbf{q}_{b},t;\mathbf{r}_{a},\mathbf{q}_{a},0)\,\varrho_{\chi}(\mathbf{r}_{a},\mathbf{q}_{a})\notag\\
	 &=\frac{1}{m^{2}}\Bigl[\mathbf{D}_{2}(t)\cdot\mathbf{A}(t)\cdot\mathbf{D}_{2}(t)\Bigr]_{lm}+\left\{\frac{\varsigma^{2}}{2}\Bigl[\mathbf{D}^{2}_{1}(t)\Bigr]_{lm}+\frac{1}{2m^{2}\varsigma^{2}}\Bigl[\mathbf{D}^{2}_{2}(t)\Bigr]_{lm}\right\}\,,\label{E:wouwkn}\\
	 \langle\,\mathbf{p}_{b}^{(l)}\mathbf{p}_{b}^{(m)}\rangle&=-\int\!d\mathbf{r}_{b}d\mathbf{q}_{b}\!\int\!d\mathbf{r}_{a}d\mathbf{q}_{a}\;\delta^{(N)}(\mathbf{q}_{b})\,\Bigl[\frac{1}{2}\,\frac{\partial}{\partial\mathbf{r}^{(l)}_{b}}+\frac{\partial}{\partial\mathbf{q}^{(l)}_{b}}\Bigr]\Bigl[\frac{1}{2}\,\frac{\partial}{\partial\mathbf{r}^{(m)}_{b}}+\frac{\partial}{\partial\mathbf{q}^{(m)}_{b}}\Bigr]\notag\\
	&\qquad\qquad\qquad\qquad\qquad\qquad\qquad\qquad\times J(\mathbf{r}_{b},\mathbf{q}_{b},t;\mathbf{r}_{a},\mathbf{q}_{a},0)\,\varrho_{\chi}(\mathbf{r}_{a},\mathbf{q}_{a};0)\notag\\
	 &=\Bigl[\dot{\mathbf{D}}_{2}(t)\cdot\mathbf{A}\cdot\dot{\mathbf{D}}_{2}(t)\Bigr]_{lm}+\mathbf{B}_{lm}+\Bigl[\dot{\mathbf{D}}_{2}(t)\cdot\mathbf{C}\Bigr]_{lm}+\Bigl[\mathbf{C}^{T}\cdot\dot{\mathbf{D}}^{T}_{2}(t)\Bigr]_{lm}\notag\\
	 &\qquad\qquad\qquad\quad+\left\{\frac{m^{2}\varsigma^{2}}{2}\,\Bigl[\dot{\mathbf{D}}_{1}(t)\dot{\mathbf{D}}^{T}_{1}(t)\Bigr]{}_{lm}+\frac{1}{2\varsigma^{2}}\,\Bigl[\dot{\mathbf{D}}_{2}(t)\dot{\mathbf{D}}^{T}_{2}(t)\Bigr]{}_{lm}\right\}\,,\\
	 \langle\,\pmb{\chi}_{b}^{(l)}\mathbf{p}_{b}^{(m)}\rangle&=-i\!\int\!d\mathbf{r}_{b}d\mathbf{q}_{b}\int\!d\mathbf{r}_{a}d\mathbf{q}_{a}\;\delta^{(N)}(\mathbf{q}_{b})\,\Bigl[\mathbf{r}_{b}^{(l)}+\frac{1}{2}\,\mathbf{q}_{b}^{(l)}\Bigr]\Bigl[\frac{1}{2}\,\frac{\partial}{\partial\mathbf{r}^{(m)}_{b}}+\frac{\partial}{\partial\mathbf{q}^{(m)}_{b}}\Bigr]\notag\\
	&\qquad\qquad\qquad\qquad\qquad\qquad\qquad\qquad\times J(\mathbf{r}_{b},\mathbf{q}_{b},t;\mathbf{r}_{a},\mathbf{q}_{a},0)\,\varrho_{\chi}(\mathbf{r}_{a},\mathbf{q}_{a})\notag\\
	 &=\frac{i}{2}\,\delta_{lm}+\frac{1}{m}\,\Bigl[\mathbf{D}_{2}(t)\cdot\mathbf{A}\cdot\dot{\mathbf{D}}_{2}(t)\Bigr]_{lm}+\frac{1}{m}\,\Bigl[\mathbf{D}_{2}(t)\cdot\mathbf{C}\Bigr]_{lm}\notag\\
	 &\qquad\qquad\qquad+\left\{\frac{m\varsigma^{2}}{2}\,\Bigl[\mathbf{D}_{1}(t)\cdot\dot{\mathbf{D}}_{1}(t)\Bigr]_{lm}+\frac{1}{2m\varsigma^{2}}\,\Bigl[\mathbf{D}_{2}(t)\cdot\dot{\mathbf{D}}_{2}(t)\Bigr]_{lm}\right\}\,,\\
	 \langle\,\mathbf{p}_{b}^{(m)}\pmb{\chi}_{b}^{(l)}\rangle&=-i\,\delta_{lm}+\langle\,\pmb{\chi}_{b}^{(l)}\mathbf{p}_{b}^{(m)}\rangle\,.\label{E:youwkn}
\end{align}
In addition we have $\langle\,\pmb{\chi}_{b}^{(l)}\,\rangle=0$ and $\langle\,\mathbf{p}_{b}^{(l)}\,\rangle=0$ due to the initial density matrix $\varrho_{\chi}(t_{a})$ we choose. Note that the commutation relation is preserved. Furthermore, each correlation in \eqref{E:wouwkn} to \eqref{E:youwkn} can be decomposed into two subgroups. The first group, which depends on $\varsigma$ in this case, is called the intrinsic or active component. Basically it is caused by the intrinsic quantum fluctuations associated with the internal degrees of freedom; thus it depends on their wavefunctions. On the other hand, the second group, called the induced or passive component, is induced from the quantum fluctuations of the environment scalar field, thus independent of the wavefunctions of the internal degrees of freedom. In general, the active components gradually decay to zero with time due to the dissipation caused by the environment. On the other hand, the passive components increases from zero owing to the quantum fluctuations of the environment. Usually it will grow without bound if we carelessly ignore the contribution of the dissipation backreaction \cite{HWL09}. The fluctuation-dissipation relation ensures the energy balance between these two processes such that the induced or passive components will approach a saturated value and remains bounded.

%At first sight, it seems that both components are quite irrelevant because they are caused by %different mechanisms, but in fact it turns out that they are tightly connected with each other by %the fluctuation-dissipation relation associated with the environment. Let us put it this way; the %fluctuating motion of the system $\chi$ may modify the environment in a way similar to the %radiation due to a moving charge,

At late time $t\to\infty$, %when the motion of the internal degrees of freedom $\chi^{(l)}$ reaches equilibrium,
when the active components vanish, the correlation functions are completely determined by the induced components,
\begin{align}
	 \langle\,\pmb{\chi}_{b}^{(l)}\pmb{\chi}_{b}^{(m)}\rangle&=\frac{g^{2}}{m^{2}}\int^{\infty}_{-\infty}\!\frac{d\omega}{2\pi}\;\Bigl[\widetilde{\pmb{\mathfrak{D}}}_{2}(\omega)\cdot\widetilde{\mathbf{G}}_{H}(\omega)\cdot\widetilde{\pmb{\mathfrak{D}}}_{2}(-\omega)\Bigr]_{lm}\,,\label{E:nkqqmna}
\intertext{and}
	 \langle\,\mathbf{p}_{b}^{(l)}\mathbf{p}_{b}^{(m)}\rangle&=g^{2}\int^{\infty}_{-\infty}\!\frac{d\omega}{2\pi}\;\omega^{2}\,\Bigl[\widetilde{\pmb{\mathfrak{D}}}_{2}(\omega)\cdot\widetilde{\mathbf{G}}_{H}(\omega)\cdot\widetilde{\pmb{\mathfrak{D}}}_{2}(-\omega)\Bigr]_{lm}\,,\label{E:rnkwia}\\
	 \frac{1}{2}\langle\,\{\pmb{\chi}_{b}^{(l)},\mathbf{p}_{b}^{(m)}\}\rangle&=\frac{i}{m}\int^{\infty}_{-\infty}\!\frac{d\omega}{2\pi}\;\omega\,\Bigl[\widetilde{\pmb{\mathfrak{D}}}_{2}(\omega)\cdot\widetilde{\mathbf{G}}_{H}(\omega)\cdot\widetilde{\pmb{\mathfrak{D}}}_{2}(-\omega)\Bigr]_{lm}=0\,,
\end{align}
where we have chosen $g_{i}=g$ and defined a new kernel function $\pmb{\mathfrak{D}}_{2}(s)=\theta(s)\mathbf{D}_{2}(s)$, and its Fourier transform $\widetilde{\pmb{\mathfrak{D}}}_{2}(\omega)$ is given by
\begin{equation}\label{E:snwnskag}
	 \widetilde{\pmb{\mathfrak{D}}}_{2}(\omega)=\left[-\omega^{2}\,\mathbf{I}+\pmb{\Omega}^{2}-\dfrac{g^{2}}{m}\,\widetilde{\mathbf{G}}_{R}(\omega)\right]^{-1}\,,
\end{equation}
if we define the Fourier transformation $\tilde{f}(\omega)$ of a given function $f(t)$ by
\begin{equation}\label{E:wjkndsaa}
	\tilde{f}(\omega)=\int_{-\infty}^{\infty}\!d\tau\;e^{i\,\omega\tau}\,f(\tau)\,.
\end{equation}
It can be further shown that the integrand of \eqref{E:nkqqmna} becomes
\begin{equation}\label{E:dfkmlssw}
	 \widetilde{\pmb{\mathfrak{D}}}_{2}(\omega)\cdot\widetilde{\mathbf{G}}_{H}(\omega)\cdot\widetilde{\pmb{\mathfrak{D}}}^{*}_{2}(\omega)=\frac{m}{g^{2}}\,\coth\frac{\beta\omega}{2}\;\operatorname{Im}\widetilde{\pmb{\mathfrak{D}}}_{2}(\omega)\,,
\end{equation}
and the correlation functions in \eqref{E:nkqqmna}--\eqref{E:rnkwia} reduce to
\begin{align}
	 \langle\,\pmb{\chi}_{b}^{(l)}\pmb{\chi}_{b}^{(m)}\rangle&=\frac{1}{m}\,\operatorname{Im}\int^{\infty}_{-\infty}\!\frac{d\omega}{2\pi}\;\coth\frac{\beta\omega}{2}\Bigl[\widetilde{\pmb{\mathfrak{D}}}_{2}(\omega)\Bigr]_{lm}\,,\label{E:kwenkjsdb}\\
	 \langle\,\mathbf{p}_{b}^{(l)}\mathbf{p}_{b}^{(m)}\rangle&=m\,\operatorname{Im}\int^{\infty}_{-\infty}\!\frac{d\omega}{2\pi}\;\omega^{2}\coth\frac{\beta\omega}{2}\Bigl[\widetilde{\pmb{\mathfrak{D}}}_{2}(\omega)\Bigr]_{lm}\,.
\end{align}
Once we know the correlations between the detectors, we may construct the covariance matrix out of them. Eqs.~\eqref{E:wouwkn}--\eqref{E:youwkn} constitutes the elements of the covariance matrix, which plays a central role in determining the entanglement of the corresponding system.

\subsection{Fluctuation-dissipation}

Since the Green's functions, in particular, the retarded Green's function $G_{R}$ and the Hadamard function $G_{H}$ play a key role in determining the backreactions of the scalar field environment and the dynamics of the reduced system, it is worthy to expound their properties. We observe that for the thermal states of the massless scalar field, the retarded Green's function is given by
\begin{align}
	 G_{R}(\mathbf{x},t;\mathbf{x}',t')&=i\,\theta(t-t')\,\operatorname{Tr}_{\phi}\Bigl(\bigl[\phi(\mathbf{x},t),\phi(\mathbf{x}',t')\bigr]\times\varrho_{\beta}\Bigr)\notag\\
	 &=i\,\theta(t-t')\int\frac{d^{3}\mathbf{k}}{(2\pi)^{3}}\frac{1}{2\omega}\,\Bigl[e^{i\mathbf{k}\cdot(\mathbf{x}-\mathbf{x}')-i\omega(t-t')}-e^{-i\mathbf{k}\cdot(\mathbf{x}-\mathbf{x}')+i\omega(t-t')}\Bigr]\notag\\
	&=\theta(\tau)\,\frac{1}{4\pi R}\Bigl[\delta(\tau-R)-\delta(\tau+R)\Bigr]=\frac{1}{2\pi }\,\theta(\tau)\,\delta(\tau^{2}-R^{2})\,,
\end{align}
with $\tau=t-t$ and $\mathbf{R}=\mathbf{x}-\mathbf{x}'$. It is independent of the temperature $\beta^{-1}$ of the environment. In the limit $R\to0$, we see that
\begin{equation*}
	\lim_{R\to0}\frac{1}{2R}\Bigl[\delta(\tau+R)-\delta(\tau-R)\Bigr]=\delta'(\tau)\,,
\end{equation*}
so, in this case, the retarded Green's function reduces to
\begin{equation}\label{E:wdnksnaz}
	\lim_{\mathbf{x}'\to\mathbf{x}}G_{R}(\mathbf{x},t;\mathbf{x}',t')=-\frac{1}{2\pi }\,\theta(\tau)\,\delta'(\tau)\,.
\end{equation}
The Fourier transform of $G_{R}(\mathbf{x},t;\mathbf{x}',t')$ can be derived from \eqref{E:wjkndsaa} as
\begin{align}
	 \widetilde{G}_{R}(\omega)&=\int_{-\infty}^{\infty}\!d\tau\;e^{i\omega\tau}G_{R}(\mathbf{x},t;\mathbf{x}',t')=\frac{1}{4\pi R}\,e^{i\omega R}\,,\label{E:wekakauqi}
\end{align}
if $\mathbf{x}$, $\mathbf{x}'$ are not explicit functions of $t$ and $t'$ respectively. Apparently it gives $\widetilde{G}_{R}^{\vphantom{*}}(-\omega)=\widetilde{G}_{R}^{*}(\omega)$. Care has to be taken for the limit $R\to0$, where $\widetilde{G}_{R}(\omega)$ has the real divergence. Usually it is interpreted as a form of self-energy and can be absorbed as part of the renormalized mass,
\begin{equation}
	\lim_{\mathbf{x}'\to\mathbf{x}}\widetilde{G}_{R}(\omega)=\text{divergence}+i\,\frac{\omega}{4\pi}\,.
\end{equation}
Eq.~\eqref{E:wdnksnaz} implies that the self-force on $\chi$, which essentially is the integral expression in \eqref{E:howswqkz}, will be effectively reduced to a local term.

The Hadamard function of the scalar field in its thermal state is given by
\begin{align}
	 G_{H,\,\beta}(\mathbf{x},t;\mathbf{x}',t')&=\frac{1}{2}\operatorname{Tr}_{\phi}\Bigl(\bigl\{\phi(\mathbf{x},t),\phi(\mathbf{x}',t')\bigr\}\times\varrho_{\beta}\Bigr)\notag\\
	 &=\sum_{n=0}^{\infty}\int\frac{d^{3}\mathbf{k}}{(2\pi)^{3}}\frac{1}{2\omega}\,\bigl(1-e^{-\beta\omega}\bigr)e^{-n\beta\omega}\Bigl(n+\frac{1}{2}\Bigr)\Bigl[e^{i\mathbf{k}\cdot\mathbf{R}-i\omega\tau}+e^{-i\mathbf{k}\cdot\mathbf{R}+i\omega\tau}\Bigr]\notag\\
	 &=\frac{1}{4\pi^{2}R}\int_{-\infty}^{\infty}d\omega\;\operatorname{sgn}(\omega)\Bigl[\frac{e^{-\beta\lvert\omega\rvert}}{1-e^{-\beta\lvert\omega\rvert}}+\frac{1}{2}\Bigr]\,\sin\omega R\,e^{-i\omega\tau}\label{E:gjkndsaa}\,,
\end{align}
again with $\tau=t-t$, $\mathbf{R}=\mathbf{x}-\mathbf{x}'$, and $\beta^{-1}$ being the temperature of the thermal state of the scalar field. Carrying out \eqref{E:gjkndsaa} directly gives
\begin{align}
	 G_{H,\,\beta}(\mathbf{x},t;\mathbf{x}',t')&=-\frac{1}{4\pi^{2}}\sum_{n=-\infty}^{\infty}\frac{1}{(\tau-i\,n\beta)^{2}-R^{2}}\\
	&=-\frac{1}{8\pi R\,\beta}\left[\coth\frac{\pi(\tau-R)}{\beta}-\coth\frac{\pi(\tau+R)}{\beta}\right]\,.
\end{align}
From \eqref{E:gjkndsaa}, we can immediately read out the Fourier transform of the Hadamard function at finite temperature $\beta^{-1}$,
\begin{equation}\label{E:uyriwnq}
	 \widetilde{G}_{H,\,\beta}(\omega)=\operatorname{sgn}(\omega)\Bigl[\frac{e^{-\beta\lvert\omega\rvert}}{1-e^{-\beta\lvert\omega\rvert}}+\frac{1}{2}\Bigr]\,\frac{\sin\omega R}{2\pi R}=\coth\frac{\beta\omega}{2}\,\frac{\sin\omega R}{4\pi R}\,.
\end{equation}
In the limit $R\to0$, it reduces to
\begin{equation}
	 \lim_{\mathbf{x}'\to\mathbf{x}}\widetilde{G}_{H,\,\beta}(\omega)=\frac{\lvert\omega\rvert}{2\pi}\,\Bigl[\frac{e^{-\beta\lvert\omega\rvert}}{1-e^{-\beta\lvert\omega\rvert}}+\frac{1}{2}\Bigr]=\frac{\omega}{4\pi}\,\coth\frac{\beta\omega}{2}\,.
\end{equation}
Compare \eqref{E:wekakauqi} with \eqref{E:uyriwnq}, we see that
\begin{equation}
	\widetilde{G}_{H,\,\beta}(\omega)=\coth\frac{\beta\omega}{2}\,\operatorname{Im}\widetilde{G}_{R}(\omega)\,.
\end{equation}
This is the fluctuation-dissipation relation.

Now we can make a connection with the $N$-detector case. There we have the retarded Green's function given by $\mathbf{G}^{ij}_{R}(\mathbf{z}^{(i)}(t),t;\mathbf{z}^{(j)}(t'),t')$, with $\mathbf{z}^{(i)}$ being the trajectory of the $i^{th}$ detector. If we assume that the trajectories of the probes never intersect, then $R=\lvert\mathbf{z}^{(i)}-\mathbf{z}^{(j)}\rvert\neq0$ for $i\neq j$. In the special case that all the probes are located at fixed positions, $\widetilde{G}_{R}(\omega)$ in \eqref{E:snwnskag} will be essentially given by \eqref{E:wekakauqi} with $R=\lvert\mathbf{z}^{(i)}-\mathbf{z}^{(j)}\rvert$. This implies that
\begin{equation}\label{E:tnwnskag}
	 \widetilde{\pmb{\mathfrak{D}}}_{2}(-\omega)=\left[-\omega^{2}\mathbf{I}+\pmb{\Omega}^{2}-\dfrac{g^{2}}{m}\,\widetilde{\mathbf{G}}_{R}(-\omega)\right]^{-1}=\left[-\omega^{2}\mathbf{I}+\pmb{\Omega}^{2}-\dfrac{g^{2}}{m}\,\widetilde{\mathbf{G}}_{R}^{*}(\omega)\right]^{-1}=\widetilde{\pmb{\mathfrak{D}}}_{2}^{*}(\omega)\,.
\end{equation}
On the other hand, we have $\widetilde{G}^{ij}_{H,\,\beta}(\omega)$ given by
\begin{equation}
	\widetilde{G}^{ij}_{H,\,\beta}(\omega)=\coth\frac{\beta\omega}{2}\,\frac{\sin\omega R}{4\pi R}\,,
\end{equation}
with $R=\lvert\mathbf{z}^{(i)}-\mathbf{z}^{(j)}\rvert$, if the probes are located at fixed positions.

%\newpage

\section{The covariance matrix and entanglement measure}
Since the covariance matrix contains the essential information about a many-body quantum system, before we delve into the entanglement structure of a $N$ oscillators system (NOS) it is helpful to spend a little space to describe its main features.  The covariance matrix is a generalization of the variance function for a single variable theory to a multiple-component theory. For illustration purpose with relevance to the subsequent discussions, let us consider an $N$-mode quantum system~\cite{SMD94} with the canonical variables $\chi^{(j)}$ and $p^{(k)}$, which satisfy the commutation relations
\begin{equation}\label{E:lrjpqs}
	 \bigl[\chi^{(j)},p^{(k)}\bigr]=i\,\delta_{jk}\,,\qquad\qquad\bigl[\chi^{(j)},\chi^{(k)}\bigr]=\bigl[p^{(j)},p^{(k)}\bigr]=0\,.
\end{equation}
It proves convenient to define a vector $\mathbf{R}$ containing $2N$ canonical variables,
\begin{equation}
	\mathbf{R}=(\chi^{(1)},p^{(1)},\dots,\chi^{(N)},p^{(N)})^{T}\,,
\end{equation}
The commutation relations corresponding to \eqref{E:lrjpqs} can then be written as
\begin{align}\label{E:erlkna}
	 \bigl[\mathbf{R},\mathbf{R}^{T}\bigr]&=i\,\mathbf{J}\,,&\mathbf{J}&=\bigoplus_{k=1}^{N}\,\pmb{\omega}\,,&\pmb{\omega}&=\begin{pmatrix}0	 &+1\\-1	&0\end{pmatrix}\,.
\end{align}
Now if we apply a real linear transformation, represented by a $2N\times2N$ real matrix $\mathbf{S}$, upon the canonical variables $\chi$ and $p$,
\begin{equation}
	\mathbf{R}\to \mathbf{R}'=\mathbf{S}\cdot\mathbf{R}\,,
\end{equation}
The condition that $\mathbf{R}'$ remain as canonical variables, namely, that $\mathbf{R}'$ obey the same commutation relations \eqref{E:erlkna} as $\mathbf{R}$,  demands that the transformation matrix $\mathbf{S}$ must satisfy
\begin{equation}\label{E:bwaamfsk}
	 \bigl[\mathbf{R}',\mathbf{R}'^{T}\bigr]=\mathbf{S}\bigl[\mathbf{R},\mathbf{R}^{T}\bigr]\mathbf{S}^{T}=i\,\mathbf{S}\cdot\mathbf{J}\cdot\mathbf{S}^{T}=i\,\mathbf{J}\,,\qquad\qquad\Rightarrow\qquad\qquad \mathbf{S}\cdot\mathbf{J}\cdot\mathbf{S}^{T}=\mathbf{J}\,.
\end{equation}
This turns out to be the definition for an element in the real symplectic group $Sp(2N,\mathbb{R})$. Thus the transformation represented by $\mathbf{S}$ is called the symplectic transformation. Since the symplectic transformation maps one set of canonical variables in the phase space to another, it enjoys particular importance in physics.

\subsection{Covariance matrix}

Let the expectation value of an observable $\pmb{\mathcal{O}}$ be given by $\langle\pmb{\mathcal{O}}\rangle=\operatorname{Tr}\bigl\{\pmb{\mathcal{O}}\,\varrho\bigr\}$, where $\varrho$ is the density matrix operator of a given state. Without loss of generality, we may assume that the expectation value of $\mathbf{R}$ in the state $\varrho$ is zero, i.e., $\langle\mathbf{R}\rangle=0$, because even if the mean value does not vanish, we can still make it happen by shifting the variable $\mathbf{R}\rightarrow\mathbf{R}-\langle\mathbf{R}\rangle$. Next consider the matrix $\mathbf{R}\,\mathbf{R}^{T}$. If we decompose its expectation value into
\begin{equation}
	\langle \mathbf{R}\,\mathbf{R}^{T}\rangle=\frac{1}{2}\,\langle\bigl\{\mathbf{R},\,\mathbf{R}^{T}\bigr\}\rangle+\frac{1}{2}\langle\bigl[\mathbf{R},\,\mathbf{R}^{T}\bigr]\rangle=\mathbf{V}+\frac{i}{2}\,\mathbf{J}\,,
\end{equation}
The matrix $\mathbf{V}$ on the right hand side is called the covariance matrix:
\begin{equation*}
	\mathbf{V}=\frac{1}{2}\,\langle\bigl\{\mathbf{R},\,\mathbf{R}^{T}\bigr\}\rangle\,.
\end{equation*}
Sometimes it proves handy to express the covariance matrix $\mathbf{V}$ by $2\times2$ block matrices
\begin{equation}
	V=\begin{pmatrix}
			\pmb{\sigma}_{11}		&\pmb{\sigma}_{12}	&\cdots	&\pmb{\sigma}_{1N}\\[6pt]
			\pmb{\sigma}_{12}^{T}	&\pmb{\sigma}_{22}	&\cdots	&\pmb{\sigma}_{2N}\\[6pt]
			\vdots					&\vdots				&\ddots	&\vdots\\[6pt]
			\pmb{\sigma}_{1N}^{T}	&\pmb{\sigma}_{2N}^{T}				&\cdots	&\pmb{\sigma}_{NN}
	  \end{pmatrix}\,,
\end{equation}
where
\begin{equation}
	\pmb{\sigma}_{jk}=\begin{pmatrix}\dfrac{1}{2}\langle\bigl\{\chi^{(j)},\chi^{(k)}\bigr\}\rangle &\dfrac{1}{2}\langle\bigl\{\chi^{(j)},p^{(k)}\bigr\}\rangle\\[12pt]
	\dfrac{1}{2}\langle\bigl\{p^{(j)},\chi^{(k)}\bigr\}\rangle &\dfrac{1}{2}\langle\bigl\{p^{(j)},p^{(k)}\bigr\}\rangle
	\end{pmatrix}\,.
\end{equation}
Note that $\pmb{\sigma}_{kk}$ is symmetric. It should be apparent from the construction that the covariance matrix is indeed a generalization of the variance  of single variable functions.

According to the Williamson theorem~\cite{JW36}, we can always find some symplectic matrix $\mathbf{S}$ that will transform the symmetric covariance matrix $\mathbf{V}$ into a diagonal form
\begin{equation}\label{E:wkbwkwa}
	\widetilde{\mathbf{V}}=\bigoplus_{k=1}^{N}\operatorname{diag}(\eta_{k},\eta_{k})\,.
\end{equation}
This implies that we can always find a new set of canonical bases such that there is no correlation between them. The parameters $\eta_{k}$ are the symplectic eigenvalues of the covariance matrix $\mathbf{V}$. Alternatively they can be found by solving the symplectic eigenvalue problem,
\begin{equation}
	\Bigl(\mathbf{V}-i\,\eta\,\mathbf{J}\Bigr)\cdot\mathbf{v}=0\,,
\end{equation}
with the corresponding eigenvector $\mathbf{v}$, or by the conventional eigenvalue problem
\begin{equation}
	\Bigl(i\,\mathbf{J}\cdot\mathbf{V}-\eta\Bigr)\cdot\mathbf{v}=0\,,
\end{equation}
for the matrix $i\,\mathbf{J}\cdot\mathbf{V}$. Obviously the symplectic eigenvalues, as well as the determinant of the covariance matrix, are invariant under the symplectic transformations. In fact, from eq.~\eqref{E:wkbwkwa}, we see that the uncertainty principle requires
\begin{equation}
	\eta_{k}^{2}\geq\frac{1}{4}\,,
\end{equation}
for all $k$ because in the basis $(\tilde{\chi},\tilde{p})$ that diagonalizes the covariance matrix $\mathbf{V}$ into the Williamson normal form $\widetilde{\mathbf{V}}$, each $2\times2$ block matrix along the diagonal has the form
\begin{equation}
	\begin{pmatrix}
		\langle\,\tilde{\chi}^{(k)\,2}\,\rangle &0\\
		0 &\langle\,\tilde{p}^{(k)\,2}\,\rangle
	\end{pmatrix}\,,
\end{equation}
and, for each mode $k$, the uncertainty principle rules that $\langle\,\tilde{\chi}^{(k)\,2}\,\rangle\langle\,\tilde{p}^{(k)\,2}\,\rangle\geq1/4$. Hence it implies that if the covariance matrix $\mathbf{V}$ is physically realizable, then its symplectic eigenvalues must be greater than or equal to $1/2$. This further shows that
\begin{equation}\label{E:mklww}
	\widetilde{\mathbf{V}}+\frac{i}{2}\,\mathbf{J}=\bigoplus_{k=1}^{N}\begin{pmatrix} \eta_{k}	 &\dfrac{i}{2}\\[4pt]-\dfrac{i}{2}	&\eta_{k}\end{pmatrix}\geq0\,.
\end{equation}
Since the covariance matrix can always be put into the diagonal form \eqref{E:wkbwkwa} by a symplectic transformation, eq.~\eqref{E:mklww} asserts that by virtue of the uncertainty principle a physically realizable covariance matrix must satisfy
\begin{equation}\label{E:ksnuwpa}
	\mathbf{V}+\frac{i}{2}\,\mathbf{J}\geq0\,,
\end{equation}
that is, it must be positive-definite.

\subsection{Entanglement: a primer}

For the benefit of the novice and for the sake of completeness we preface the discussion of entanglement measure by a short summary of what quantum entanglement is.  In the pure state cases, if a state $\lvert\Psi_{s}\rangle$ can be written as a product of pure states such as
\begin{equation}\label{E:fnkswuea}
	\lvert\Psi_{s}\rangle=\lvert\psi_{1}\rangle\otimes\lvert\psi_{2}\rangle\,,
\end{equation}
for $\lvert\psi_{1}\rangle\in\mathcal{H}_{1}$ and $\lvert\psi_{2}\rangle\in\mathcal{H}_{2}$ in a bipartite system, then it is called a product state or a separable state. If the state $\lvert\Psi_{e}\rangle$ cannot be written as \eqref{E:fnkswuea}, then $\lvert\Psi_{e}\rangle$ is an entangled state. A famous example of an entangled state for a bipartite system is the Bell state
\begin{equation}
	 \lvert\Psi_{b}\rangle=\frac{1}{\sqrt{2}}\Bigl(\lvert\uparrow\uparrow\rangle+\lvert\downarrow\downarrow\rangle\Bigr)\,,
\end{equation}
where $\lvert\uparrow\rangle$, $\lvert\downarrow\rangle$ are two possible outcomes of a qubit state.

Here comes the natural question:  If we are given a pure state $\lvert\Psi\rangle\in\mathcal{H}=\mathcal{H}_{1}\otimes\mathcal{H}_{2}$,
\begin{equation}\label{E:qlaml}
	\lvert\Psi\rangle=\sum_{i,\,j}d_{ij}\,\lvert\psi_{i}\rangle\otimes\lvert\phi_{j}\rangle\,.
\end{equation}
with $\lvert\psi_{i}\rangle\in\mathcal{H}_{1}$ and $\lvert\phi_{j}\rangle\in\mathcal{H}_{2}$, how do we know whether the state $\lvert\Psi\rangle$ can be written into a product state like \eqref{E:fnkswuea} with an appropriate choice of bases? The answer is provided by the singular value decomposition. It says that for every complex matrix $\mathfrak{D}$, there always exist unitary transformations $\mathfrak{U}$ and $\mathfrak{V}$ such that $\mathfrak{U}\cdot\mathfrak{D}\cdot\mathfrak{V}$ is diagonal with real, non-negative diagonal elements. Thus for each state $\lvert\Psi\rangle$, we can always find suitable bases $\lvert\psi_{i}^{s}\rangle$ and $\lvert\phi_{i}^{s}\rangle$ in terms of which \eqref{E:qlaml} reduces to
\begin{equation}
	\lvert\Psi\rangle=\sum_{i}\sqrt{\lambda_{i}}\;\lvert\psi_{i}^{s}\rangle\otimes\lvert\phi_{i}^{s}\rangle\,,
\end{equation}
where the $\lambda_{i}$ are known as Schmidt coefficients, and the upper limit of the sum is determined by the dimension of the smaller subsystem.

It is then straightforward to see that, compared with \eqref{E:fnkswuea}, if there is only one nonvanishing Schmidt coefficient, then the state $\lvert\Psi\rangle$ is a separable state, and vice versa. Since the singular values $\lambda_{i}$ are uniquely defined, the number of Schmidt coefficients provides a convenient criterion of separability of a state.

In realistic situations we more often have to deal with mixed states, such as, when we consider an open quantum system, which interacts with an environment. Due to its huge number of degrees of freedom we cannot  track all the details of the motion of the environment. If what we need is the overall effect of the environment on the system we can trace out the environment in a fashion similar to \eqref{E:woejs} and obtain a reduced description for the state of the open system which is usually mixed.

For the mixed state cases, the density matrix of a quantum system takes on the generic form
\begin{equation}
	\varrho=\sum_{k}p_{k}\lvert\psi_{k}\rangle\langle\psi_{k}\rvert\,,
\end{equation}
where $\lvert\psi_{k}\rangle$ is a set of complete bases for the system. The weight $p_{k}$ is the probability of findng the system in the state $\lvert\psi_{k}\rangle$. Unitarity requires that
\begin{equation}
	\sum_{k}p_{k}=1\,.
\end{equation}
A well-known simple criterion for discerning a mixed state from a pure one is: Given the density matrix of the state $\varrho$, if the state is pure, the trace of $\varrho^{2}$ is equal to 1. One the other hand, if it is a mixed state, then the trace is less than 1, that is,
\begin{align}
	&\text{pure state}&\sum_{k}p_{k}^{2}&=1\,,\\
	&\text{mixed state}&\sum_{k}p_{k}^{2}&<1\,.
\end{align}

The simplest mixed state for a bipartite system is described by the so-called mixed product state
\begin{equation}
	\varrho=\varrho^{(1)}\otimes\varrho^{(2)}
\end{equation}
where $\varrho^{(1)}$ and $\varrho^{(2)}$ are the density matrices of the respective subsystems. More generally, the joint system can be described by a convex sum of different product states
\begin{equation}\label{E:koalma}
	\varrho=\sum_{i}p_{i}\,\varrho_{i}^{(1)}\otimes\varrho_{i}^{(2)}
\end{equation}
with $p_{i}>0$ and $\displaystyle\sum_{i}p_{i}=1$. The corresponding state is called the separable mixed state. From this definition, if a mixed state of a bipartite system cannot be expressed as \eqref{E:koalma}, then the state is entangled. In other words, a mixed state $\varrho$ is entangled if there are no states $\varrho^{(1)}_{i}\in\mathcal{H}_{1}$, $\varrho_{i}^{(2)}\in\mathcal{H}_{2}$, and non-negative value of $p_{i}$, such that $\varrho$ can be expressed as a convex sum like \eqref{E:koalma}.

It has bee shown~\cite{AP96,H396} that if a mixed state is separable, then the partial transpose of the density matrix remains positive
\begin{equation}\label{E:loalma}
	\varrho^{pt_{1}}=\sum_{i}p_{i}\,\varrho_{i}^{(1)T}\otimes\varrho_{i}^{(2)}>0\,.
\end{equation}
where $T$ in the superscript represents the transpose operation. Here, as an example, we perform transposition on only $\varrho^{(1)}$. This is the positive-partial-transpose (PPT) separability criterion. Thus, in principle, if we find a negative eigenvalue for the partially transposed density matrix, then the corresponding state is entangled. The sufficient condition of the separability criterion, on the other hand, is valid only for low-dimensional bipartite systems~\cite{H396}. This powerful criterion is further extended to an $M\times N$ bi-partitions of the $(M+N)$-mode bi-symmetric Gaussian system~\cite{SAI05}. Both sufficient and necessary conditions apply to such a Gaussian system.

When we deal with a system described by continuous variables, the criterion based on the density matrix of the system becomes inconvenient to work with because the dimension of the density matrix for continuous variables is usually infinite. This is where the covariance matrix finds its use.

\subsection{Entanglement measure}

When translated into the covariance matrix, the partial transpose of the density matrix is equivalent to the mirror reflection of the canonical momentum in the corresponding subsystem.  Let the mirror reflection operator be $\pmb{\Gamma}$. The Peres-Horodecki positive partial transpose criterion~\cite{RS00} then says
\begin{equation}\label{E:gwusa}
	 \bar{\mathbf{V}}+\frac{i}{2}\,\mathbf{J}\geq0\,,\qquad\qquad\qquad\bar{\mathbf{V}}=\pmb{\Gamma}\cdot\mathbf{V}\cdot\pmb{\Gamma}^{T}\,,
\end{equation}
because when $\varrho^{pt}$ is a bona fide density matrix, $\bar{\mathbf{V}}$ will be a bona fide covariance matrix, and then eq.~\eqref{E:ksnuwpa} implies eq.~\eqref{E:gwusa}. Conversely, the entanglement criterion~\cite{AF07} then says that if the state is entangled, the partial transpose of its density matrix is negative, or equivalently the symplectic values $\bar{\eta}$ of the corresponding partially transposed covariance matrix can be smaller than $1/2$.

A simple and computable entanglement measure known as `negativity' is defined based on the idea that if one of the eigenvalues $\lambda_{k}$ of $\varrho^{pt}$ is negative, then $\varrho$ is entangled~\cite{AP96, H396}. Thus the negativity $\mathcal{N}$ is given by~\cite{VW02}
\begin{equation}\label{E:ernkott}
	\mathcal{N}=\frac{1}{2}\sum_{k}\Bigl(\lvert\lambda_{k}\rvert-\lambda_{k}\Bigr)\,.
\end{equation}
If $\varrho^{pt}$ is positive semi-definite, then the negativity $\mathcal{N}$ vanishes. On the other hand it will take on positive values if $\varrho^{pt}$ has at least one negative eigenvalue. However, due to the limitation of the positive-partial-transpose (PPT) criterion~\cite{H396}, the negativity is fully valid only for low-dimensional systems such as a $2\times2$ or a $2\times3$ system.

An alternative entanglement measure, which is often used for the continuous-variable system, is a variant of the logarithmic negativity $E_{\mathcal{N}}$~\cite{AF07,MP05}. It basically collects the symplectic eigenvalues of the partially-transposed covariance matrix, which are smaller than $1/2$,
\begin{equation}\label{E:yueow}
	E_{\mathcal{N}}=\sum_{k}\max\Bigl\{0,-\ln2\bar{\eta}_{k}\Bigr\}\geq0\,.
\end{equation}
The equal sign can be satisfied by the ppt entangled state~\cite{H396} because in higher dimensions there are entangled states which have a positive partial-transposed density matrix operator.

%\newpage

%%%%%%%%%%%%%%%%%%%%%%%

\section{Disparate Inter-detector Couplings for $N=3$}\label{S:rnfekjna}

Now we are in a position to discuss entanglement between three strongly coupled detectors. Let these detectors be labeled by $Q$, $A$, $B$. Assume that the coupling strength $\lambda$ between the $QA$ pair is the same as that between the $QB$ pair, but different from the coupling strength $\sigma$ between $A$ and $B$. This leads to an interaction matrix $\pmb{\Omega}^{2}$ of the form
\begin{equation}\label{E:eriekns}
	\pmb{\Omega}^{2}=\begin{pmatrix}
				\omega_{0}^{2}+2\lambda	&-\lambda	&-\lambda\\[6pt]
				-\lambda					&\omega_{0}^{2}+\lambda+\sigma	&-\sigma\\[6pt]
				-\lambda		&-\sigma		&\omega_{0}^{2}+\lambda+\sigma
				\end{pmatrix}\,.
\end{equation}
It describes the case of disparate coupling between three detectors, that is, $\sigma_{12}=\sigma_{13}=\lambda$ but $\sigma_{23}=\sigma$ in \eqref{E:jeaiekjna}. These coupling strengths are not assumed weak in comparison with the parameter $\omega_{0}^{2}$, which is the oscillator frequency of the internal degrees of freedom (or the renormalized oscillating frequency after absorbing the divergence in the retarded Green's function) in the absence of the inter-detector coupling. Nonetheless, we still assume the  coupling $g$ between the detectors and the environmental scalar field is weak.

In general, the motion of such a system is highly non-Markovian due to the multi-time correlations generated in the sharing of a common bath. This makes it difficult to  find the normal modes of the motion. However, for certain configurations, this can be done. To accomplish this we first  discuss how to do the diagonalization of the interaction matrix.

The eigenvalues $\nu_{i}$ and the normalized eigenvectors $\mathbf{v}_{i}$ for the interaction matrix $\pmb{\Omega}^{2}$ are given by
\begin{align}
	\nu_{1}&=\omega_{0}^{2}\,,&\mathbf{v}_{1}^{T}&=\frac{1}{\sqrt{3}}\,(1,1,1)\,,\label{E:nkerww}\\
	\nu_{2}&=\omega_{0}^{2}+3\lambda\,,&\mathbf{v}_{3}^{T}&=\frac{1}{\sqrt{6}}\,(2,-1,-1)\,,\\
	\nu_{3}&=\omega_{0}^{2}+\lambda+2\sigma\,,&\mathbf{v}_{3}^{T}&=\frac{1}{\sqrt{2}}\,(0,1,-1)\,,\label{E:okerww}
\end{align}
If we construct the matrix $\mathbf{U}$ in terms of the normalized eigenvectors, it will transform the interaction matrix $\pmb{\Omega}^{2}$ into the diagonal form $\pmb{\Lambda}^{2}$ by $\mathbf{U}^{T}\cdot\pmb{\Omega}^{2}\cdot\mathbf{U}$,
\begin{equation}\label{E:huwnak}
	\mathbf{U}=\begin{pmatrix} \dfrac{1}{\sqrt{3}} &\dfrac{2}{\sqrt{6}} &0\\[10pt]\dfrac{1}{\sqrt{3}}	 &-\dfrac{1}{\sqrt{6}}	&\dfrac{1}{\sqrt{2}}\\[10pt]
	\dfrac{1}{\sqrt{3}}	&-\dfrac{1}{\sqrt{6}} &-\dfrac{1}{\sqrt{2}}
	\end{pmatrix}\,,\qquad\qquad\qquad\pmb{\Lambda}^{2}=\begin{pmatrix} \omega_{0}^{2} &0	&0\\[16pt]0	 &\omega_{0}^{2}+3\lambda	&0\\[16pt]0	&0	&\omega_{0}^{2}+\lambda+2\sigma\end{pmatrix}\,.
\end{equation}
However, the elements of the $\pmb{\Omega}^{2}$ matrix has the property that
\begin{equation}\label{E:bkssw}
	\sum_{i}\left[\pmb{\Omega}^{2}\right]_{ij}=\sum_{j}\left[\pmb{\Omega}^{2}\right]_{ij}=\text{const.}\,,
\end{equation}
which is a consequence of the way we construct the Lagrangian for inter-detector coupling. It has an interesting implication. Suppose we try to solve the eigenvalue problem for such a system in general. Let the eigenvector $\mathbf{v}$ be given by $\mathbf{v}=(a_{1},a_{2},a_{3})^{T}$,  we need to solve a simultaneous set of homogeneous equations
\begin{equation}\label{E:jkdhw}
	\pmb{\Omega}^{2}\cdot\mathbf{v}=\nu\,\mathbf{v}\,,\qquad\Rightarrow\qquad\begin{cases}
	 \bigl(\left[\pmb{\Omega}^{2}\right]_{11}-\nu\bigr)\,a_{1}+\left[\pmb{\Omega}^{2}\right]_{12}\,a_{2}+\left[\pmb{\Omega}^{2}\right]_{13}\,a_{3}=0\,,\\
	 \left[\pmb{\Omega}^{2}\right]_{21}\,a_{1}+\bigl(\left[\pmb{\Omega}^{2}\right]_{22}-\nu\bigr)a_{2}+\left[\pmb{\Omega}^{2}\right]_{23}\,a_{3}=0\,,\\
	 \left[\pmb{\Omega}^{2}\right]_{31}\,a_{1}+\left[\pmb{\Omega}^{2}\right]_{32}\,a_{2}+\bigl(\left[\pmb{\Omega}^{2}\right]_{33}-\nu\bigr)a_{3}=0\,.
	\end{cases}
\end{equation}
Adding these three equations together leads to
\begin{equation}\label{E:oeruohnw}
	\bigl(\sum_{i}\left[\pmb{\Omega}^{2}\right]_{ij}-\nu\bigr)\bigl(a_{1}+a_{2}+a_{3}\bigr)=0\,.
\end{equation}
Thus we have either $\sum_{i}\left[\pmb{\Omega}^{2}\right]_{ij}-\nu=0$ or $a_{1}+a_{2}+a_{3}=0$. This implies that i) one of the eigenvalues must be equal to the sum of elements in one of the rows or columns of the matrix $\pmb{\Omega}^{2}$, and that ii) the elements of the eigenvectors for the rest of the eigenvalues must sum to zero.

From the condition i), we see from the first equation of \eqref{E:jkdhw} that the elements of the corresponding eigenvector must be such that $a_{1}=a_{2}=a_{3}$, so we have
\begin{equation}\label{E:reeottie}
	\nu_{1}=\sum_{i}\left[\pmb{\Omega}^{2}\right]_{ij}\,,\qquad\qquad\mathbf{v}_{1}^{T}=\frac{1}{\sqrt{3}}\,(1,1,1)\,,
\end{equation}
after proper normalization. The second condition states that the rest of the eigenvector must be normal to the plane $a_{1}+a_{2}+a_{3}=0$. Therefore as long as the matrix $\pmb{\Omega}^{2}$ satisfies \eqref{E:bkssw}, the matrix $\mathbf{U}$ that diagonalizes $\pmb{\Omega}^{2}$ will be the same up to re-ordering of the rows or columns. If we define the normal modes $\pmb{\mathfrak{v}}$ by $\pmb{\mathfrak{v}}=\mathbf{U}^{T}\cdot\mathbf{r}$, then it implies one of the normal modes, say $\pmb{\mathfrak{v}}^{(1)}$, must be such that
\begin{equation}
	 \pmb{\mathfrak{v}}^{(1)}=\mathbf{v}_{1}^{T}\cdot\mathbf{r}=\frac{\mathbf{r}^{(1)}+\mathbf{r}^{(2)}+\mathbf{r}^{(3)}}{\sqrt{3}}\,,
\end{equation}
which highlights the special role of the center-of-mass coordinate of the original variables of motion. Extension to the case of $N$ detectors is straightforward. This seemingly intuitive yet rarely proven fact will play a key role in our analysis of one important facet of macroscopic quantum phenomena later.

Now in terms of normal mode variables, the two-point function matrix $\mathbf{G}$, shown in \eqref{E:rehkwd} for example, will be transformed into
\begin{equation}
	\pmb{\mathfrak{G}}=\mathbf{U}^{T}\cdot\mathbf{G}\cdot\mathbf{U}\,,
\end{equation}
and it applies to both the retarded Green's function $\mathbf{G}_{R}$ or the Hadamard function $\mathbf{G}_{H}$. Let us examine the structure of the transformed Green's function. Generically, we have
\begin{align}
	 \pmb{\mathfrak{G}}_{11}&=\mathbf{G}_{11}+\frac{2}{3}\Bigl(\mathbf{G}_{12}+\mathbf{G}_{13}+\mathbf{G}_{23}\Bigr)\,,\\
	\pmb{\mathfrak{G}}_{12}&=\frac{1}{3\sqrt{2}}\,\Bigl(\mathbf{G}_{12}+\mathbf{G}_{13}-2\mathbf{G}_{23}\Bigr)\,,\\
	\pmb{\mathfrak{G}}_{13}&=\frac{1}{\sqrt{6}}\,\Bigl(\mathbf{G}_{12}-\mathbf{G}_{13}\Bigr)\,,\\
	\pmb{\mathfrak{G}}_{23}&=\frac{1}{\sqrt{3}}\,\Bigl(\mathbf{G}_{12}-\mathbf{G}_{13}\Bigr)\,,\\
	 \pmb{\mathfrak{G}}_{22}&=\mathbf{G}_{11}+\frac{1}{3}\Bigl(-2\mathbf{G}_{12}-2\mathbf{G}_{13}+\mathbf{G}_{23}\Bigr)\,,\\
	\pmb{\mathfrak{G}}_{33}&=\mathbf{G}_{11}-\mathbf{G}_{23}\,.
\end{align}
due to the fact that $\mathbf{G}_{11}=\mathbf{G}_{22}=\mathbf{G}_{33}$. If the coupling constant between the oscillator and the environment is the same for all three detectors, and if the distances between any two of the detectors are the same, then we will have $\mathbf{G}_{ij}$, $i\neq j$ all the same. That is, if the detectors sit at the vertices of an equilateral triangle, the Green function matrix, which describes the backreaction effects from the environment, will take on only two distinct values.

Let $\mathbf{G}_{11}=\mathbf{G}_{22}=\mathbf{G}_{33}=\mathbf{G}^{\odot}$ and $\mathbf{G}_{ij}=\mathbf{G}^{\Pi}$ for $i\neq j$. For future reference, we write down the explicit expressions of $\mathbf{G}^{\odot}_{R}$ and $\mathbf{G}^{\Pi}_{R}$,
\begin{align}
	\mathbf{G}^{\odot}_{R}(\mathbf{x},t;,\mathbf{x},t')&=-\frac{1}{2\pi}\,\theta(\tau)\,\delta'(\tau)\,,\\
	\mathbf{G}^{\Pi}_{R}(\mathbf{x},t;\mathbf{x}',t')&=\frac{1}{2\pi}\,\theta(t-t')\,\delta(\tau^{2}-d^{2})\,,
\end{align}
where $\tau=t-t'$ and $d=\lvert\mathbf{x}-\mathbf{x}'\rvert$, for the retarded Green's functions. Furthermore, we immediately see the consequence of this particular configuration leads to $\pmb{\mathfrak{G}}_{ij}=0$ for $i\neq j$, but
\begin{align}
	\pmb{\mathfrak{G}}_{11}&=\mathbf{G}^{\odot}+2\mathbf{G}^{\Pi}\,,\\
	\pmb{\mathfrak{G}}_{22}&=\mathbf{G}^{\odot}-\mathbf{G}^{\Pi}\,,\\
	\pmb{\mathfrak{G}}_{33}&=\mathbf{G}^{\odot}-\mathbf{G}^{\Pi}\,.
\end{align}
In the condensed notation, we will write them as $\mathbf{g}_{ii}=\mathbf{G}^{\odot}+\alpha_{i}\,\mathbf{G}^{\Pi}$, where $\alpha_{1}=2$, $\alpha_{2}=\alpha_{3}=-1$. The transformed Green's function matrix $\mathbf{g}$ becomes diagonal, too. It implies that we will have three decoupled equations for the normal modes $\pmb{\mathfrak{v}}$,
\begin{align}
	 \ddot{\mathfrak{v}}_{1}(t)+2\gamma\,\dot{\mathfrak{v}}_{1}(t)+\omega_{0}^{2}\,\mathfrak{v}_{1}(t)-\frac{4\gamma}{d}\,\mathfrak{v}_{1}(t-d)&=0\,,\\
	 \ddot{\mathfrak{v}}_{2}(t)+2\gamma\,\dot{\mathfrak{v}}_{2}(t)+\bigl(\omega_{0}^{2}+3\lambda\bigr)\,\mathfrak{v}_{2}(t)+\frac{2\gamma}{d}\,\mathfrak{v}_{2}(t-d)&=0\,,\\
	 \ddot{\mathfrak{v}}_{3}(t)+2\gamma\,\dot{\mathfrak{v}}_{3}(t)+\bigl(\omega_{0}^{2}+\lambda+2\sigma\bigr)\,\mathfrak{v}_{3}(t)+\frac{2\gamma}{d}\,\mathfrak{v}_{3}(t-d)&=0\,.
\end{align}
As long as the coupling between the detectors and the environment is sufficiently weak and the distance between detectors is not too short, the time-delay term, which describes the mutual indirect influence between oscillators, only introduces a small correction for the late-time dynamics~\cite{LinHu09, HWL08}. Hence the fundamental solutions to the above equations look like those to the damped oscillators, and are given by
\begin{equation}\label{E:fheuye}
	 \mathfrak{d}_{1}^{(i)}(s)=e^{-\Gamma_{i}s}\Bigl[\cos\Omega_{i}s+\frac{\Gamma_{i}}{\Omega_{i}}\,\sin\Omega_{i}s\Bigr]\,,\qquad\mathfrak{d}_{2}^{(i)}(s)=e^{-\Gamma_{i}s}\,\frac{1}{\Omega_{i}}\,\sin\Omega_{i}s\,,
\end{equation}
where the damping constant $\Gamma_{i}$ and the ``resonance'' frequency\footnote{Strictly speaking this does not look like the resonance frequency. The typical resonance frequency has a contribution of the order $\gamma^{2}$, which is ignored here by the assumption that $1/\omega_{i}d>>\gamma$. However we should keep in mind that this assumption is not necessary.} $\Omega_{i}$ are given by
\begin{equation}\label{E:wkenwkaaq}
	 \Gamma_{i}=\gamma\left[1+\frac{\alpha_{i}}{\omega_{i}d}\,\sin\omega_{i}d\right]\,,\qquad\qquad\Omega_{i}=\omega_{i}\left[1-\alpha_{i}\,\frac{\gamma}{\omega_{i}^{2}d}\,\cos\omega_{i}d\right]\,.
\end{equation}
in terms of the (renormalized) oscillator frequency $\omega_{i}$
\begin{equation}
	 \omega_{1}^{2}=\omega_{0}^{2}\,,\qquad\qquad\omega_{2}^{2}=\omega_{0}^{2}+3\lambda\,,\qquad\qquad\omega_{3}^{2}=\omega_{0}^{2}+\lambda+2\sigma\,,
\end{equation}
and the parameter $\gamma=g^{2}/8\pi m$.

With the choice of the initial density matrix for $\chi$ given in \eqref{E:erjdkjnds}, the density matrix still takes the same form even in terms of the normal mode variables owing  to the fact that the transformation matrix $\mathbf{U}$ is orthogonal. Let the normal mode variables that correspond to $\chi$ be denoted by $\zeta$. Then it can be shown that the correlations between $\chi$ and their conjugate variables can be expressed in terms of the counterparts of the normal modes $\zeta$, that is,
\begin{align}
	 \langle\,\pmb{\chi}_{b}^{(l)}\pmb{\chi}_{b}^{(m)}\rangle&=\mathbf{U}_{li}\,\mathbf{U}_{mj}\,\langle\,\pmb{\zeta}_{b}^{(i)}\pmb{\zeta}_{b}^{(j)}\rangle\,,\\
	 \langle\,\mathbf{p}_{b}^{(l)}\mathbf{p}_{b}^{(m)}\rangle&=\mathbf{U}_{li}\,\mathbf{U}_{mj}\,\langle\,\pmb{\pi}_{b}^{(i)}\pmb{\pi}_{b}^{(j)}\rangle\,,\\
	 \langle\,\pmb{\chi}_{b}^{(l)}\mathbf{p}_{b}^{(m)}\rangle&=\mathbf{U}_{li}\,\mathbf{U}_{mj}\,\langle\,\pmb{\zeta}_{b}^{(i)}\pmb{\pi}_{b}^{(j)}\rangle\,,
\end{align}
where $\pi$ is the canonical momentum conjugate to $\zeta$. More importantly the correlation functions for the normal modes have the same form as those for $\chi$ in \eqref{E:wouwkn}--\eqref{E:youwkn} except that all the relevant matrices are replaced by their counterparts associated with the normal modes. This correspondence greatly simplifies the calculation when the normal modes can be completely decoupled, as shown in the configuration introduced previously.
Hence, let us turn our attention to the evolution of the normal mode variables. Since each normal mode $\pmb{\zeta}^{(i)}$ behaves like a damped oscillators driven by the quantum fluctuations of the environment scalar field, it can be shown that the active component of each correlation function, say, $\langle\,\pmb{\zeta}_{b}^{(i)}{}^{2}\rangle$, decays to zero as time evolves due to dissipation as a form of backreaction, but the passive component, on the other hand, grows from zero, due to the environment noise, gradually to a saturated value. This, as we remarked earlier, is a consequence of the energy balance between dissipation and fluctuations~\cite{HWL08}. This value is also independent of the initial state of the normal mode.

Owing to the fact that the motion of the normal modes are totally decoupled, the matrices $\pmb{\mathscr{D}}_{2}=\mathbf{U}^{T}\cdot\mathbf{D}_{2}\cdot\mathbf{U}$ and $\pmb{\mathfrak{G}}_{H}$ are diagonalized. Thus at late-time $t\to\infty$, the only nonzero terms left are the passive components like $\langle\,\pmb{\zeta}^{(i)\,2}_{b}\rangle$ and $\langle\,\pmb{\pi}^{(i)\,2}_{b}\rangle$. As seen from \eqref{E:kwenkjsdb}, they are
\begin{align}
	 \langle\,\pmb{\zeta}^{(i)\,2}_{b}\rangle&=\frac{1}{m}\operatorname{Im}\int^{\infty}_{-\infty}\!\frac{d\omega}{2\pi}\;\widetilde{\mathfrak{d}}_{2}^{(i)}(\omega)\,,\label{E:anwkj}\\
	 \langle\,\pmb{\pi}^{(i)\,2}_{b}\rangle&=m\,\operatorname{Im}\int^{\infty}_{-\infty}\!\frac{d\omega}{2\pi}\;\omega^{2}\,\widetilde{\mathfrak{d}}_{2}^{(i)}(\omega)\,,\label{E:bnwkj}\\
	\langle\,\{\pmb{\zeta}^{(i)}_{b},\,\pmb{\pi}_{b}^{(i)}\}\rangle&=0\,,
\end{align}
where $\widetilde{\mathfrak{d}}_{2}^{(i)}$ is the Fourier transformation of the fundamental solution \eqref{E:fheuye} to each normal mode, that is
\begin{equation}
	\widetilde{\pmb{\mathscr{D}}}_{2}=\begin{pmatrix} \widetilde{\mathfrak{d}}_{2}^{(1)} &0	&0\\
	0	&\widetilde{\mathfrak{d}}_{2}^{(2)}	&0\\
	0	&0	&\widetilde{\mathfrak{d}}_{2}^{(3)}\end{pmatrix}\,,
\end{equation}
and
\begin{equation}
	 \widetilde{\mathfrak{d}}_{2}^{(i)}(\omega)=\left[-\omega^{2}+\omega_{i}^{2}-\frac{g^{2}}{m}\,\widetilde{\mathbf{g}}_{ii}(\omega)\right]^{-1}=\left[-\omega^{2}+\omega_{i}^{2}-i\,2\gamma\,\omega-2\gamma\, \frac{\alpha_{i}}{d}\,e^{i\,\omega d}\right]^{-1}\,,
\end{equation}
and we have used
\begin{equation}
	 \widetilde{\mathbf{G}}^{\odot}(\omega)=\text{div.}+i\,\frac{\omega}{4\pi}\,,\qquad\qquad\widetilde{\mathbf{G}}^{\Pi}(\omega)=\frac{1}{4\pi d}\,e^{i\,\omega d}\,.
\end{equation}
The divergence in $\widetilde{\mathbf{G}}^{\odot}(\omega)$ can be absorbed into $\omega_{i}$. Here we have assumed that we have a zero-temperature bath, so that $\beta\to\infty$ and $\coth(\beta\omega/2)=1$.

Carrying out the integrals in \eqref{E:anwkj} and \eqref{E:bnwkj} gives
\begin{align}
	 \langle\,\pmb{\zeta}^{(i)\,2}_{b}\rangle&=\frac{1}{2m\Omega_{i}}\left[1-\frac{2}{\pi}\frac{\Gamma_{i}}{\Omega_{i}}+\cdots\right]\notag\\
	 &=\frac{1}{2m\omega_{i}}\left\{1-\frac{2}{\pi}\frac{\gamma_{i}}{\omega_{i}}+\frac{\gamma\,\alpha_{i}}{\omega_{i}^{2}d}\biggl[\cos\omega_{i}d-\frac{2}{\pi}\,\sin\omega_{i}d\biggr]+\cdots\right\}\,,\\
	 \langle\,\pmb{\pi}^{(i)\,2}_{b}\rangle&=\frac{m\Omega_{i}}{2}\left[1+\frac{2}{\pi}\frac{\Gamma_{i}}{\Omega_{i}}\left(\ln\frac{\Pi^{2}}{\Omega_{i}^{2}}-1\right)+\cdots\right]\notag\\
	 &=\frac{m\omega_{i}}{2}\left\{1+\frac{2}{\pi}\frac{\gamma_{i}}{\omega_{i}}\,\beta_{i}-\frac{\gamma\,\alpha_{i}}{\omega_{i}^{2}d}\biggl[\cos\omega_{i}d-\frac{2}{\pi}\,\beta_{i}\,\sin\omega_{i}d\biggr]+\cdots\right\}\,,
\end{align}
with
\begin{equation*}
	\beta_{i}=\ln\frac{\Pi^{2}}{\omega_{i}^{2}}-1\,.
\end{equation*}
where $\Pi$ is the cutoff frequency associate with the scalar field, and assumed to take the same value for all oscillators. Note that the correction due to interaction with the environment is linear in $\gamma$. The damping constant $\Gamma_{i}$ and the resonance frequency $\Omega_{i}$ are given by \eqref{E:wkenwkaaq}
\begin{equation*}
	 \Gamma_{i}=\gamma\left[1+\frac{\alpha_{i}}{\omega_{i}d}\,\sin\omega_{i}d\right]\,,\qquad\qquad\Omega_{i}=\omega_{i}\left[1-\alpha_{i}\,\frac{\gamma}{\omega_{i}^{2}d}\,\cos\omega_{i}d\right]\,.
\end{equation*}
with the parameter $\gamma=g^{2}/8\pi m$. In the weak coupling limit $\gamma\ll\omega_{i}$, the resonance frequency $\Omega_{i}$ also differs from the normalized oscillating frequency $\omega_{i}$ by the order of $\gamma/\omega_{i}$. Thus to leading order we have
\begin{align}
	\langle\,\pmb{\zeta}^{(i)\,2}_{b}\rangle&=\begin{cases}
											\dfrac{\varsigma^{2}}{2}\,,&t\to0\,,\\[12pt]
											\dfrac{1}{2m\omega_{i}}\,,&t\to\infty\,.
										\end{cases}
\intertext{and}
	\langle\,\pmb{\pi}^{(i)\,2}_{b}\rangle&=\begin{cases}
											\dfrac{1}{2\varsigma^{2}}\,,&t\to0\,,\\[12pt]
											\dfrac{m\omega_{i}}{2}\,,&t\to\infty\,.
										\end{cases}
\end{align}

Next we can transform the results for the normal modes back into the counterparts for the original variables $\pmb{\chi}$ by
\begin{align}
	 \langle\,\pmb{\chi}_{b}^{(l)}\pmb{\chi}_{b}^{(m)}\rangle&=U_{li}\,U_{mj}\,\langle\,\pmb{\zeta}_{b}^{(i)}\pmb{\zeta}_{b}^{(j)}\rangle\,,&&\Rightarrow&\langle\,\pmb{\chi}_{b}^{\vphantom{T}}\,\pmb{\chi}_{b}^{T}\rangle&=\mathbf{U}\,\langle\,\pmb{\zeta}_{b}^{\vphantom{T}}\,\pmb{\zeta}_{b}^{T}\rangle\,\mathbf{U}^{T}\,,\\
	 \langle\,\mathbf{p}_{b}^{(l)}\mathbf{p}_{b}^{(m)}\rangle&=U_{li}\,U_{mj}\,\langle\,\pmb{\pi}_{b}^{(i)}\pmb{\pi}_{b}^{(j)}\rangle\,,&&\Rightarrow&\langle\,\mathbf{p}_{b}^{\vphantom{T}}\,\mathbf{p}_{b}^{T}\rangle&=\mathbf{U}\,\langle\,\pmb{\pi}_{b}^{\vphantom{T}}\,\pmb{\pi}_{b}^{T}\rangle\,\mathbf{U}^{T}\,.
\end{align}
Explicitly, we have $\langle\,\pmb{\chi}_{b}^{(l)}\pmb{\chi}_{b}^{(m)}\rangle$ given by
\begin{align}
	 \langle\,\pmb{\chi}_{b}^{(1)\,2}\,\rangle&=\frac{1}{3}\Bigl[\langle\,\pmb{\zeta}_{b}^{(1)\,2}\,\rangle+2\langle\,\pmb{\zeta}_{b}^{(2)\,2}\,\rangle\Bigr]\,,\label{E:oeowjx}\\
	 \langle\,\pmb{\chi}_{b}^{(2)\,2}\,\rangle=\langle\,\pmb{\chi}_{b}^{(3)\,2}\,\rangle&=\frac{1}{6}\Bigl[2\langle\,\pmb{\zeta}_{b}^{(1)\,2}\,\rangle+\langle\,\pmb{\zeta}_{b}^{(2)\,2}\,\rangle+3\langle\,\pmb{\zeta}_{b}^{(3)\,2}\,\rangle\Bigr]\,,\\
	 \langle\,\pmb{\chi}_{b}^{(1)}\pmb{\chi}_{b}^{(2)}\,\rangle=\langle\,\pmb{\chi}_{b}^{(1)}\pmb{\chi}_{b}^{(3)}\,\rangle&=\frac{1}{3}\Bigl[\langle\,\pmb{\zeta}_{b}^{(1)\,2}\,\rangle-\langle\,\pmb{\zeta}_{b}^{(2)\,2}\,\rangle\Bigr]\,,\label{E:krnktiw}\\
	 \langle\,\pmb{\chi}_{b}^{(2)}\pmb{\chi}_{b}^{(3)}\,\rangle&=\frac{1}{6}\Bigl[2\langle\,\pmb{\zeta}_{b}^{(1)\,2}\,\rangle+\langle\,\pmb{\zeta}_{b}^{(2)\,2}\,\rangle-3\langle\,\pmb{\zeta}_{b}^{(3)\,2}\,\rangle\Bigr]\,,\label{E:neowjx}
\end{align}
and similar structures for $\langle\,\mathbf{p}_{b}^{(l)}\mathbf{p}_{b}^{(m)}\rangle$. Also note that in these cases we have $\langle\,\pmb{\chi}^{(i)}_{b}\pmb{\chi}^{(j)}_{b}\,\rangle=\langle\,\pmb{\chi}^{(j)}_{b}\pmb{\chi}^{(i)}_{b}\,\rangle$, and thus
\begin{equation}
	 \frac{1}{2}\,\langle\,\{\pmb{\chi}^{(i)}_{b},\pmb{\chi}^{(j)}_{b}\}\,\rangle=\langle\,\pmb{\chi}^{(i)}_{b}\pmb{\chi}^{(j)}_{b}\,\rangle\,.
\end{equation}

Now it is instructive to make comparison with the case that there is no inter-detector coupling~\cite{NOS1}. In the latter case, the late-time correlation between the variables $\pmb{\chi}^{(i)}$ solely results from their interaction with the environment, thus it is typically of the order $\gamma$. In addition, as a consequence of the intervention of the background field, this correlation depends on the spatial separation between the detectors. Thus the detectors tend to have the pairwise correlation due to the facts that they have stronger correlation if they get closer to one another, and that additional correlation mediated by the third parties will be at least of the order $\mathcal{O}(\gamma^{2})$. On the other hand, in the presence of inter-detector coupling, the later-time correlation comes from the superposition of their counterparts from the normal modes; it is of the same order as the uncertainty of the corresponding variables. Thus the correlation due to inter-detector coupling tends to be much stronger than the induced correlation by the shared bath, and it does not depend on the distance between two detectors. These are the fundamental differences between the two cases studied in the present two papers.

\subsection{Bi-partite entanglement}

In the subsequent discussion, since the contributions from the environment are much smaller than those from inter-detector coupling, we will only keep terms which are zeroth order in $\gamma$. The covariance matrix at late time is then explicitly given by
\begin{equation}\label{E:hyeiw}
	\pmb{\sigma}=\begin{pmatrix}
			\langle\,\pmb{\chi}_{b}^{(1)}\pmb{\chi}_{b}^{(1)}\,\rangle &0 &\langle\,\pmb{\chi}_{b}^{(1)}\pmb{\chi}_{b}^{(2)}\,\rangle &0 &\langle\,\pmb{\chi}_{b}^{(1)}\pmb{\chi}_{b}^{(3)}\,\rangle &0\\
			0 &\langle\,\mathbf{p}_{b}^{(1)}\mathbf{p}_{b}^{(1)}\,\rangle &0 &\langle\,\mathbf{p}_{b}^{(1)}\mathbf{p}_{b}^{(2)}\,\rangle &0 &\langle\,\mathbf{p}_{b}^{(1)}\mathbf{p}_{b}^{(3)}\,\rangle\\
			\langle\,\pmb{\chi}_{b}^{(2)}\pmb{\chi}_{b}^{(1)}\,\rangle &0 &\langle\,\pmb{\chi}_{b}^{(2)}\pmb{\chi}_{b}^{(2)}\,\rangle &0 &\langle\,\pmb{\chi}_{b}^{(2)}\pmb{\chi}_{b}^{(3)}\,\rangle &0\\
			0 &\langle\,\mathbf{p}_{b}^{(2)}\mathbf{p}_{b}^{(1)}\,\rangle &0 &\langle\,\mathbf{p}_{b}^{(2)}\mathbf{p}_{b}^{(2)}\,\rangle &0 &\langle\,\mathbf{p}_{b}^{(2)}\mathbf{p}_{b}^{(3)}\,\rangle\\
			\langle\,\pmb{\chi}_{b}^{(3)}\pmb{\chi}_{b}^{(1)}\,\rangle &0 &\langle\,\pmb{\chi}_{b}^{(3)}\pmb{\chi}_{b}^{(2)}\,\rangle &0 &\langle\,\pmb{\chi}_{b}^{(3)}\pmb{\chi}_{b}^{(3)}\,\rangle &0\\
			0 &\langle\,\mathbf{p}_{b}^{(3)}\mathbf{p}_{b}^{(1)}\,\rangle &0 &\langle\,\mathbf{p}_{b}^{(3)}\mathbf{p}_{b}^{(2)}\,\rangle &0 &\langle\,\mathbf{p}_{b}^{(3)}\mathbf{p}_{b}^{(3)}\,\rangle
		\end{pmatrix}\,,
\end{equation}
with $\langle\,\pmb{\chi}_{b}^{(i)}\pmb{\chi}_{b}^{(j)}\,\rangle$ given by \eqref{E:oeowjx}--\eqref{E:neowjx} and
\begin{equation*}
	 \langle\,\pmb{\zeta}^{(i)\,2}_{b}\rangle=\frac{1}{2m\omega_{i}}\,,\qquad\qquad\qquad\langle\,\pmb{\pi}^{(i)\,2}_{b}\rangle=\frac{m\omega_{i}}{2}\,.
\end{equation*}
Some information of entanglement for this tripartite system can be revealed in the symplectic eigenvalues of a covariance matrix which corresponds to the partially transposed density matrix of this system.

For a system which involves more than two parties, there is more than one way to take the partial transpose of the density matrix. Denote detector 1 as $Q$, and the remaining two as $A$ and $B$. We see two distinct cases, one where swapping  $A$ and $B$ leads to identical results, because Q is coupled with equal strength $\lambda$ to $A$ and $B$. This is the symmetric case we called Case [SYM] before. In the other case we called Case [ASM] before  the entanglement between $A$ and $Q$ is different from that between  $A$ and $B$ because the coupling strengths of these pairs are different, being $\lambda$ in the former and $\sigma$ in the latter. Thus there are two ways to perform the partial transpose. We can either do it to $Q$, which will be called the case $Q$ versus $AB$. Or we partially transpose variables associated with $A$, which will be called $A$ versus $QB$. The consequence of taking partial transpose in the density matrix is equivalent to changing the sign of the conjugate momentum of the corresponding subsystem. Take the case $Q$ versus $AB$ for example, the covariance matrix associated with the partial transpose of the subsystem $Q$ becomes
\begin{equation}
	\pmb{\sigma}^{pt_{Q}}=\begin{pmatrix}
			\langle\,\pmb{\chi}_{b}^{(1)}\pmb{\chi}_{b}^{(1)}\,\rangle &0 &\langle\,\pmb{\chi}_{b}^{(1)}\pmb{\chi}_{b}^{(2)}\,\rangle &0 &\langle\,\pmb{\chi}_{b}^{(1)}\pmb{\chi}_{b}^{(3)}\,\rangle &0\\
			0 &\langle\,\mathbf{p}_{b}^{(1)}\mathbf{p}_{b}^{(1)}\,\rangle &0 &-\langle\,\mathbf{p}_{b}^{(1)}\mathbf{p}_{b}^{(2)}\,\rangle &0 &-\langle\,\mathbf{p}_{b}^{(1)}\mathbf{p}_{b}^{(3)}\,\rangle\\
			\langle\,\pmb{\chi}_{b}^{(2)}\pmb{\chi}_{b}^{(1)}\,\rangle &0 &\langle\,\pmb{\chi}_{b}^{(2)}\pmb{\chi}_{b}^{(2)}\,\rangle &0 &\langle\,\pmb{\chi}_{b}^{(2)}\pmb{\chi}_{b}^{(3)}\,\rangle &0\\
			0 &-\langle\,\mathbf{p}_{b}^{(2)}\mathbf{p}_{b}^{(1)}\,\rangle &0 &\langle\,\mathbf{p}_{b}^{(2)}\mathbf{p}_{b}^{(2)}\,\rangle &0 &\langle\,\mathbf{p}_{b}^{(2)}\mathbf{p}_{b}^{(3)}\,\rangle\\
			\langle\,\pmb{\chi}_{b}^{(3)}\pmb{\chi}_{b}^{(1)}\,\rangle &0 &\langle\,\pmb{\chi}_{b}^{(3)}\pmb{\chi}_{b}^{(2)}\,\rangle &0 &\langle\,\pmb{\chi}_{b}^{(3)}\pmb{\chi}_{b}^{(3)}\,\rangle &0\\
			0 &-\langle\,\mathbf{p}_{b}^{(3)}\mathbf{p}_{b}^{(1)}\,\rangle &0 &\langle\,\mathbf{p}_{b}^{(3)}\mathbf{p}_{b}^{(2)}\,\rangle &0 &\langle\,\mathbf{p}_{b}^{(3)}\mathbf{p}_{b}^{(3)}\,\rangle
		\end{pmatrix}\,,
\end{equation}
and likewise for the case $A$ versus $QB$, it is
\begin{equation}
	\pmb{\sigma}^{pt_{A}}=\begin{pmatrix}
			\langle\,\pmb{\chi}_{b}^{(1)}\pmb{\chi}_{b}^{(1)}\,\rangle &0 &\langle\,\pmb{\chi}_{b}^{(1)}\pmb{\chi}_{b}^{(2)}\,\rangle &0 &\langle\,\pmb{\chi}_{b}^{(1)}\pmb{\chi}_{b}^{(3)}\,\rangle &0\\
			0 &\langle\,\mathbf{p}_{b}^{(1)}\mathbf{p}_{b}^{(1)}\,\rangle &0 &-\langle\,\mathbf{p}_{b}^{(1)}\mathbf{p}_{b}^{(2)}\,\rangle &0 &\langle\,\mathbf{p}_{b}^{(1)}\mathbf{p}_{b}^{(3)}\,\rangle\\
			\langle\,\pmb{\chi}_{b}^{(2)}\pmb{\chi}_{b}^{(1)}\,\rangle &0 &\langle\,\pmb{\chi}_{b}^{(2)}\pmb{\chi}_{b}^{(2)}\,\rangle &0 &\langle\,\pmb{\chi}_{b}^{(2)}\pmb{\chi}_{b}^{(3)}\,\rangle &0\\
			0 &-\langle\,\mathbf{p}_{b}^{(2)}\mathbf{p}_{b}^{(1)}\,\rangle &0 &\langle\,\mathbf{p}_{b}^{(2)}\mathbf{p}_{b}^{(2)}\,\rangle &0 &-\langle\,\mathbf{p}_{b}^{(2)}\mathbf{p}_{b}^{(3)}\,\rangle\\
			\langle\,\pmb{\chi}_{b}^{(3)}\pmb{\chi}_{b}^{(1)}\,\rangle &0 &\langle\,\pmb{\chi}_{b}^{(3)}\pmb{\chi}_{b}^{(2)}\,\rangle &0 &\langle\,\pmb{\chi}_{b}^{(3)}\pmb{\chi}_{b}^{(3)}\,\rangle &0\\
			0 &\langle\,\mathbf{p}_{b}^{(3)}\mathbf{p}_{b}^{(1)}\,\rangle &0 &-\langle\,\mathbf{p}_{b}^{(3)}\mathbf{p}_{b}^{(2)}\,\rangle &0 &\langle\,\mathbf{p}_{b}^{(3)}\mathbf{p}_{b}^{(3)}\,\rangle
		\end{pmatrix}\,.
\end{equation}
The symplectic eigenvalue $\eta$ of the covariance matrix can be found by solving the eigenvalue problem of the form
\begin{equation}
	\pmb{\sigma}\cdot\mathbf{v}=i\,\eta\,\mathbf{J}\cdot\mathbf{v}\,,\qquad\text{or equivalently}\qquad i\bigl(\mathbf{J}\cdot\pmb{\sigma}\bigr)\cdot\mathbf{v}=\eta\,\mathbf{v}\,,
\end{equation}
where the matrix $\pmb{\Sigma}$ is the fundamental symplectic matrix
\begin{equation}
	\mathbf{J}=\bigoplus_{k=1}^{N}\begin{pmatrix} 0 &1\\-1 &0\end{pmatrix}\,,
\end{equation}
and in our case $N=3$. The vector $\mathbf{v}$ is the corresponding eigenvector. The eigenvalues always appear in pairs and this is most transparent to see in the Williamson normal form of the covariance matrix
\begin{equation}
	\bigoplus_{k=1}^{n}\begin{pmatrix} \eta_{k} &0\\0 &\eta_{k}\end{pmatrix}\,,
\end{equation}
which is the diagonal form of a symmetric matrix under suitable symplectic transformations. The eigenvalues in our case are then given by the roots to the polynomial
\begin{equation}
	\det\bigl(\pmb{\sigma}^{pt}-i\,\eta\,\mathbf{J}\bigr)=0\,,
\end{equation}
Since it is a third-degree polynomial in $\eta^{2}$, its roots can always be found exactly.

%%%%%%%%%%%%%%%%%%

%@BL June 2. JT:   I have read through Sec 4.  Plse feel free to revise everything below, hopefully before the end of Sunday so I can on monday go over this part again in order to revise Sec. 5

%%%%%%%%%%%%%%%%%%%%%%%

\subsection{Entanglement of $Q$ with $AB$}

For the $Q$ versus $AB$ or what we called the symmetric [SYM] case, the symplectic eigenvalues are
\begin{align}
	\eta&=\frac{1}{2}\,,\\
	 \eta_{\pm}&=\left\{\frac{4\omega_{1}^{2}+\omega_{1}\omega_{2}+4\omega_{2}^{2}\pm2\sqrt{2}(\omega_{2}-\omega_{1})\sqrt{(2\omega_{1}+\omega_{2})(\omega_{1}+2\omega_{2})}}{36\omega_{1}\omega_{2}}\right\}^{\frac{1}{2}}\,.
\end{align}
They do not depend on $\omega_{3}$, thus independent of the coupling constant $\sigma$ between $A$ and $B$. Among these three eigenvalues, we observe that $\eta_{+}$ is always greater than $1/2$, but $\eta_{-}$ is always smaller than $1/2$, so it signals the presence of entanglement between $Q$ and the pair $AB$ according to \eqref{E:yueow}. Hence we only focus on the symplectic eigenvalue $\eta_{-}$. In terms of the coupling constant $\lambda$ between $Q$ and $A$ or $Q$ and $B$, we can show
\begin{equation}
	\eta_{-}=\begin{cases}
				\dfrac{1}{2}-\dfrac{\lambda}{2\sqrt{2}\,\omega_{0}^{2}}+\cdots\,,&\lambda\ll\omega_{0}^{2}\,,\\[12pt]
				\dfrac{3^{3/4}}{4\sqrt{2}}\,\left(\dfrac{\omega_{0}^{2}}{\lambda}\right)^{1/4}+\cdots\,,	 &\lambda\gg\omega_{0}^{2}\,,		
			\end{cases}
\end{equation}
so it ranges from $1/2$ to 0.

The fact that $\eta_{-}$ does not depend on the interaction between $A$ and $B$ may not be that surprising because the interaction between the $QA$ pair is of the same strength as the $QB$ pair, whatever happens between $A$ and $B$ will be ``equally'' distributed to $Q$ over two channels through the $QA$, $QB$ couplings. This observation is supported by the fact that the correlation functions between the canonical variables associated with $Q$ and $A$, say $\langle\,\pmb{\chi}_{b}^{(1)}\pmb{\chi}_{b}^{(2)}\,\rangle$, taking on the same value as the corresponding correlation between $Q$ and $B$, as seen in \eqref{E:krnktiw}. The value does not depend on the coupling constant between $A$ and $B$. This balance or symmetry is built-in in the parity between $A$ and $B$ since they start from the same initial configurations, have the same initial oscillating frequency, and are coupled to $Q$ with equal strength. It can be easily disrupted if their motion is out of phase by changing any of the above-mentioned factors or by  skewing their relative positions, such that the dependence on $\sigma$ will re-appear.

When we put back the influence of the environment, the conclusion that $\eta_{-}$ does not depend on $\sigma$ still holds because both $A$ and $B$ experience the same self-force, and the same non-Markovian effects mediated by the environment. In comparison, the correlations, say $\langle\,\pmb{\chi}_{b}^{(2)}\pmb{\chi}_{b}^{(3)}\,\rangle$, between $A$ and $B$, do depend on the coupling constant $\sigma$ between $A$ and $B$.

Following this line of reasoning, we see that when the coupling between $Q$ and $A$ (or $B$) is vanishing small, the entanglement of $Q$ with $AB$ gradually disappears. This is consistent with our intuition because the correlation %@\footnote{it is not clear at this point whether the correlation is of quantum nature or classical nature (i.e. due to LOCC)?}
between $Q$ and $A$ (or $B$) vanishes, as seen in \eqref{E:krnktiw}. In this limit, the only connection between $Q$ and $A$ (or $B$) comes from their coupling to the environment which is weak. If we even ignore that part, then $Q$ is essentially isolated from $A$ and $B$. Thus if $Q$ gets disentangled with $AB$, it will remain disentangled throughout its evolution. On the other hand, if $Q$ interacts strongly with  $A$ , their momenta tend to be strongly anti-correlated while their positions remain positively correlated. At the same time the symplectic eigenvalue $\eta_{-}$ deviates further away from $1/2$ from below, signaling stronger entanglement between $Q$ and $AB$ in the sense of negativity.
%@ It is not clear so far whether both observations have any connection.

\subsection{Entanglement of $A$  with $QB$}

In contrast to the case $Q$ versus~$AB$, we now turn to the case $A$ versus ~$QB$. The corresponding symplectic eigenvalues share several identical features as in the case $Q$ versus $AB$. Among the three distinct symplectic eigenvalues, one of them is always greater than $1/2$, the second is equal to $1/2$, and the third one, denoted by $\eta_{-}$, is always smaller than $1/2$, regardless of the values of $\omega_{0}^{2}$, $\lambda$ and $\sigma$,
\begin{align}
	\eta&=\frac{1}{2}\,,\\
	 \eta_{\pm}&=\frac{1}{72\omega_{1}\omega_{2}\omega_{3}}\biggl\{2\omega_{1}^{2}\bigl(3\omega_{2}+\omega_{3}\bigr)+2\omega_{2}\omega_{3}\bigl(\omega_{2}+3\omega_{3}\bigr)+\omega_{1}\bigl(3\omega_{2}^{2}-4\omega_{2}\omega_{3}+3\omega_{3}^{2}\bigr)\biggr.\notag\\
	 &\pm\biggl.\sqrt{-324\,\omega_{1}^{2}\omega_{2}^{2}\omega_{3}^{2}+\Bigl[3\omega_{1}\omega_{2}\bigl(2\omega_{1}+\omega_{1}\bigr)+2\bigl(\omega_{1}-\omega_{2}\bigr)^{2}\omega_{3}+3\bigl(\omega_{1}+2\omega_{2}\bigr)\omega_{3}^{2}\Bigr]^{2}}\biggr\}
\end{align}
Nonetheless, in this case $\eta_{-}$ does depend on the coupling constant $\sigma$ between $A$ and $B$. Generally speaking, for fixed values of $\omega_{0}^{2}$ and $\lambda$, the symplectic eigenvalue $\eta_{-}$ monotonically decreases with larger values of $\sigma$. Therefore, it implies that when the coupling between $A$ and $B$ is stronger than the interaction between $Q$ and $A$, the entanglement for $A$ versus $QB$ is stronger than the counterpart for $Q$ with $AB$, and vice versa. Thus the role inter-detector coupling plays on determining the entanglement structure is more transparent from this comparison. On the other hand, for fixed values of $\omega_{0}^{2}$ and $\sigma$, the symplectic eigenvalue $\eta_{-}$ is not always a monotonically decreasing function of $\lambda$. For small $\lambda$, the value of $\eta_{-}$ can increase with $\lambda$ to a maxima and then monotonically decreases. It is particularly significant for larger values of $\sigma$.

To better understand the behavior of $\eta_{-}$, let us take a look at some limiting cases:
%@ \footnote{trying to understand entanglement in terms of correlation and connectedness, but am not sure how far it may work...}.
First, suppose $\lambda$ is vanishingly small and $\sigma$ is fixed in value, and then the only channel for possible entanglement of $A$ with $QB$ comes from the interaction between $A$ and $B$. $Q$ is out of reach of $A$. In this case, we can (wait for the system reaches a steady state and) examine  the cross-correlation between them. We see
\begin{align}
	 \langle\,\pmb{\chi}_{b}^{(Q)}\pmb{\chi}_{b}^{(A)}\,\rangle&=\langle\,\pmb{\chi}_{b}^{(Q)}\pmb{\chi}_{b}^{(B)}\,\rangle=\frac{1}{6m}\left(\frac{1}{\omega_{1}}-\frac{1}{\omega_{2}}\right)=0\,,\label{E:nntri}\\
	 \langle\,\mathbf{p}_{b}^{(Q)}\mathbf{p}_{b}^{(A)}\,\rangle&=\langle\,\mathbf{p}_{b}^{(Q)}\mathbf{p}_{b}^{(B)}\,\rangle=\frac{m}{6}\left(\omega_{1}-\omega_{2}\right)=0\,,\label{E:ontri}
\end{align}
but
\begin{align}
	 \langle\,\pmb{\chi}_{b}^{(A)}\pmb{\chi}_{b}^{(B)}\,\rangle&=\frac{1}{4m}\left(\frac{1}{\omega_{1}}-\frac{1}{\omega_{3}}\right)>0\,,\\
	\langle\,\mathbf{p}_{b}^{(A)}\mathbf{p}_{b}^{(B)}\,\rangle&=\frac{m}{4}\left(\omega_{1}-\omega_{3}\right)<0\,,
\end{align}
due to the fact that $\omega_{3}>\omega_{2}=\omega_{1}$ in this limit.

Then we let $\lambda$ increase from zero, there comes an extra channel between $A$ and $QB$ owing to the interaction between $A$ and $Q$. In addition, $B$ can also be related to $A$ via $Q$, thus improving the connection between $A$ and $QB$. For $\lambda\neq0$, we have $\omega_{3}>\omega_{2}>\omega_{1}$ if $\sigma$ is still greater than $\lambda$, so it implies that the cross-correlations then become:
\begin{align}
	 \langle\,\pmb{\chi}_{b}^{(Q)}\pmb{\chi}_{b}^{(A)}\,\rangle&=\langle\,\pmb{\chi}_{b}^{(Q)}\pmb{\chi}_{b}^{(B)}\,\rangle=\frac{1}{6m}\left(\frac{1}{\omega_{1}}-\frac{1}{\omega_{2}}\right)>0\,,\\
	 \langle\,\mathbf{p}_{b}^{(Q)}\mathbf{p}_{b}^{(A)}\,\rangle&=\langle\,\mathbf{p}_{b}^{(Q)}\mathbf{p}_{b}^{(B)}\,\rangle=\frac{m}{6}\left(\omega_{1}-\omega_{2}\right)<0\,,
\end{align}
while
\begin{align}
	 \langle\,\pmb{\chi}_{b}^{(A)}\pmb{\chi}_{b}^{(B)}\,\rangle&=\frac{1}{12m}\left[2\left(\frac{1}{\omega_{1}}-\frac{1}{\omega_{3}}\right)+\left(\frac{1}{\omega_{2}}-\frac{1}{\omega_{3}}\right)\right]>0\,,\\
	 \langle\,\mathbf{p}_{b}^{(A)}\mathbf{p}_{b}^{(B)}\,\rangle&=\frac{m}{12}\left[2\left(\omega_{1}-\omega_{3}\right)+\left(\omega_{2}-\omega_{3}\right)\right]<0\,.
\end{align}
As $\lambda$ increases beyond $\sigma$, even we have $\omega_{2}>\omega_{3}>\omega_{1}$,  the cross-correlation between $A$ and $B$ does not change qualitatively. However, we find both $\langle\,\pmb{\chi}_{b}^{(A)}\pmb{\chi}_{b}^{(B)}\,\rangle$ and $\langle\,\mathbf{p}_{b}^{(A)}\mathbf{p}_{b}^{(B)}\,\rangle$ reach their extremal values at $\lambda=\sigma$. That is, $\langle\,\pmb{\chi}_{b}^{(A)}\pmb{\chi}_{b}^{(B)}\,\rangle$ decreases from some positive value as $\lambda$ moves away from zero, until it reaches its minimum at $\lambda=\sigma$. After that point the correlation between $\pmb{\chi}_{b}^{(A)}$ and $\pmb{\chi}_{b}^{(B)}$ monotonically increases with $\lambda$, approaching the value $1/(3m\omega_{1})$. The minimum is
\begin{equation}
	 \min\langle\,\pmb{\chi}_{b}^{(A)}\pmb{\chi}_{b}^{(B)}\,\rangle=\frac{1}{3m}\left(\frac{1}{\omega_{1}}-\frac{1}{\omega_{2}}\right)>0\,.
\end{equation}
On the other hand, the correlation $\langle\,\mathbf{p}_{b}^{(A)}\mathbf{p}_{b}^{(B)}\,\rangle$ increases from some negative values with $\lambda$ until it reaches its maximum value
\begin{equation}
	\min\langle\,\mathbf{p}_{b}^{(A)}\mathbf{p}_{b}^{(B)}\,\rangle=\frac{m}{3}\left(\omega_{1}-\omega_{2}\right)<0\,,
\end{equation}
at $\lambda=\sigma$. Beyond that point, it decreases monotonically without bound. This  partially explains the non-monotonic behavior of $\eta_{-}$ for a fixed $\sigma$ because some other factors which enter in determining the covariance matrix, such as $\langle\,\pmb{\chi}_{b}^{(Q)}\pmb{\chi}_{b}^{(A)}\,\rangle$, do not depend on $\sigma$. %@ What is more puzzling is how to understand physically the behavior of the cross-correlation between $A$ and $B$? What causes an extrema at $\lambda=\sigma$?

\begin{figure}
\centering
    \scalebox{0.8}{\includegraphics{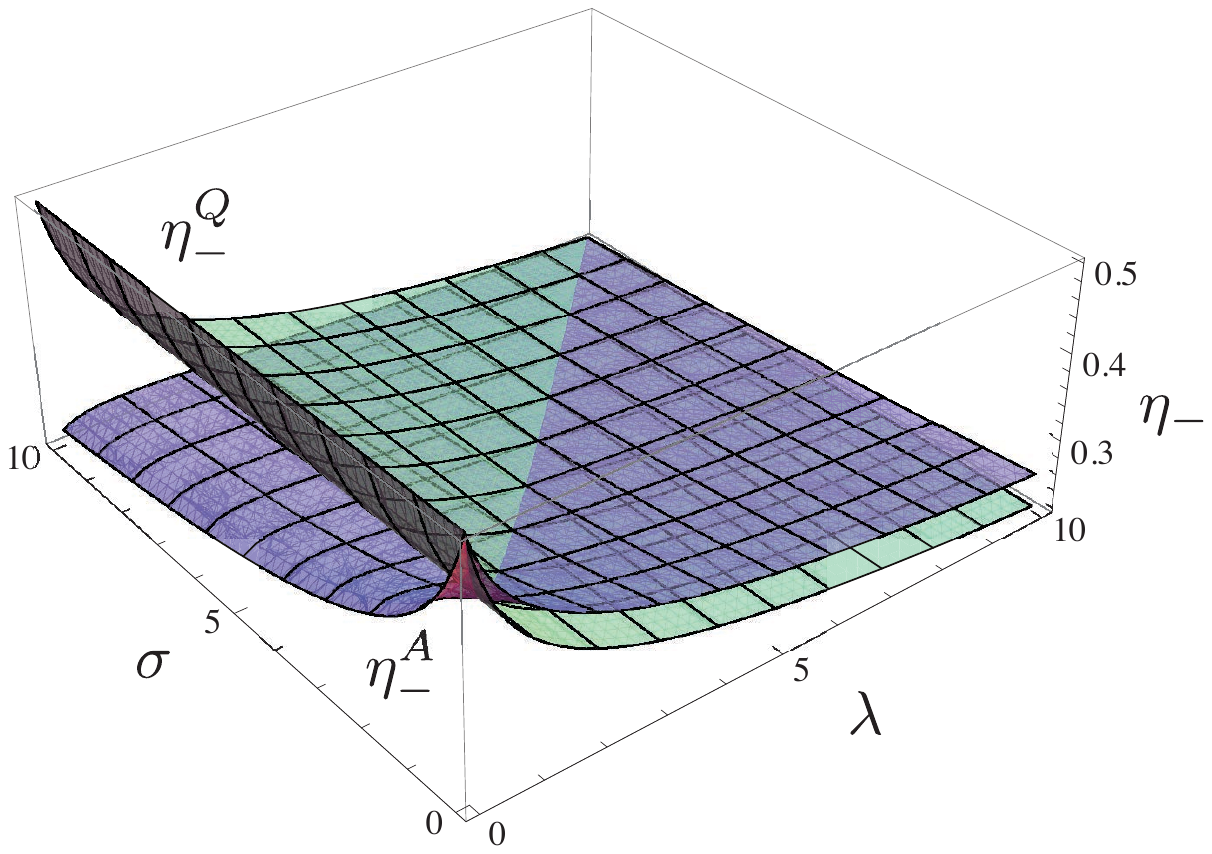}}
    \caption{Comparison of entanglement for the two cases:  $Q$ vs~$AB$ and $A$ vs~$QB$. here we show the dependence of their smallest symplectic eigenvalues, denoted by $\eta^{Q}_{-}$, $\eta^{A}_{-}$ respectively, on the inter-oscillator coupling strength $\lambda$, $\sigma$, which are scaled with respect to $\kappa$.}\label{Fi:qab}
\end{figure}

For the case with a fixed $\lambda$, if we let $\sigma=0$, that is, no direct coupling between $A$ and $B$, the only connection between $A$ and $QB$ results from interaction between $A$ and $Q$. In addition, in this case $B$ is not entirely out of contact with $A$. Their correlation can be mediated by $Q$. Now we examine their cross-correlation. For $\sigma=0$, we have $\omega_{2}>\omega_{3}>\omega_{1}$, and
\begin{align}
	 \langle\,\pmb{\chi}_{b}^{(Q)}\pmb{\chi}_{b}^{(A)}\,\rangle&=\langle\,\pmb{\chi}_{b}^{(Q)}\pmb{\chi}_{b}^{(B)}\,\rangle=\frac{1}{6m}\left(\frac{1}{\omega_{1}}-\frac{1}{\omega_{2}}\right)>0\,,\\
	 \langle\,\mathbf{p}_{b}^{(Q)}\mathbf{p}_{b}^{(A)}\,\rangle&=\langle\,\mathbf{p}_{b}^{(Q)}\mathbf{p}_{b}^{(B)}\,\rangle=\frac{m}{6}\left(\omega_{1}-\omega_{2}\right)<0\,.
\end{align}
These are in contrast with \eqref{E:nntri} and \eqref{E:ontri}. As for the cross-correlation between $A$ and $B$, we have
\begin{align}
	\langle\,\pmb{\chi}_{b}^{(A)}\pmb{\chi}_{b}^{(B)}\,\rangle&=\frac{1}{12m}\left[2\left(\frac{1} {\omega_{1}}-\frac{1}{\omega_{3}}\right)+\left(\frac{1}{\omega_{2}}-\frac{1}{\omega_{3}}\right)\right]>0\,,\\
	 \langle\,\mathbf{p}_{b}^{(A)}\mathbf{p}_{b}^{(B)}\,\rangle&=\frac{m}{12}\left[2\left(\omega_{1}-\omega_{3}\right)+\left(\omega_{2}-\omega_{3}\right)\right]<0\,.
\end{align}
These qualitative features do not change with increasing $\sigma$. However, there is a major distinction in this case. There is no extremum for the cross-correlation between $A$ and $B$ when we vary $\sigma$. This is consistent with the behavior of $\eta_{-}$. Since we have known that the entanglement $Q$ with $AB$ does not depend on $\sigma$, it is equivalent to the special case $\sigma=\lambda$. Therefore we can draw the conclusion that when the coupling $\sigma$ is greater than $\lambda$, the entanglement for the partition $A$ versus $QB$ is stronger than the counterpart for $Q$ versus $AB$. On the other hand, when the coupling $\sigma$ is weaker than $\lambda$, the entanglement for the partition $Q$ versus $AB$ is stronger than the counterpart for $A$ versus $QB$ (see Fig.~\ref{Fi:qab}).

Here we outline some general features of cross-correlations. It has been shown that with stronger inter-oscillator couplings, the cross-correlation $\langle\,\pmb{\chi}^{(i)}\pmb{\chi}^{(j)}\,\rangle$ tends to be more strongly correlated with the exception that $\langle\,\pmb{\chi}^{(Q)}\pmb{\chi}^{(A)}\,\rangle$ is not sensitive to the coupling strength $\sigma$ between $A$ and $B$. On the other hand, $\langle\,\mathbf{p}^{(i)}\mathbf{p}^{(j)}\,\rangle$ tends to be more strongly anti-correlated with stronger inter-oscillator couplings. We also observe that since $\langle\,\pmb{\chi}^{(i)}\pmb{\chi}^{(j)}\,\rangle$ is bounded, they do not change appreciably with $\lambda$ and $\sigma$. Contrary to $\langle\,\pmb{\chi}^{(i)}\pmb{\chi}^{(j)}\,\rangle$, $\langle\,\mathbf{p}^{(i)}\mathbf{p}^{(j)}\,\rangle$ is not bounded below, it can vary more dramatically with the coupling constants $\lambda$ and $\sigma$. In particular, $\langle\,\mathbf{p}^{(A)}\mathbf{p}^{(B)}\,\rangle$ varies more significantly with $\sigma$ but less with $\lambda$, while $\langle\,\mathbf{p}^{(Q)}\mathbf{p}^{(A)}\,\rangle$ changes more significantly with $\lambda$ but is independent of $\sigma$.

In conclusion, we see from this detailed analysis how the entanglement structure of coupled quantum oscillators depends on the coupling strength,  from a knowledge of the cross-correlations between the oscillators. This quantitative description is valuable in aiding our understanding of entanglement behavior because it is the simplest continuous variable many-body system which is amenable to such a detailed analysis. For larger systems with more components and less symmetry this rapidly becomes impossible. We shall  explore the implications of our results for some key issues in macroscopic quantum phenomena in the next section.

%\newpage

\section{Summary and Discussions}

\subsection{Summary of results}

The key findings are given in the form of an executive summary below (as a retribution for the length of prior sections):

\begin{itemize}

\item Please extract from the text for the main results on the influence functional, the reduced density matrix evolutionary operator, the Langevin equations, the Green's functions, the correlation functions and the covariance matrix.

\item Comparison with the case studied in \cite{NOS1} where there is no direct coupling amongst the $N$ oscillators but only field-induced coupling: With direct inter-oscillator coupling studied here, the effect of the system-environment coupling becomes of secondary importance.

\item As a measure of entanglement for the $N$ oscillator system with direct coupling studied here, it is decided by whether any of the symplectic eigenvalues of the partially transposed covariance matrix is smaller than 1/2.

\item Entanglement is enhanced with stronger coupling between any two oscillators, which confirms intuitive reasoning.

\item Two cases are studied in details where three oscillators are placed at the vertices of an equilateral triangle. Main features:

\noindent CASE [SYM] Q vs AB, where Q is coupled to A and B with equal strength:   Entanglement of Q with AB as a group is independent of the coupling between A and B.  It is as if Q is  entangled with a `center-of-mass' variable of the two oscillators AB as a group.

\noindent CASE [ASY] A vs QB, where the coupling strength between A and Q is different from between A and B.  Entanglement for A vs QB does depends on QA coupling and AB coupling.
		\begin{enumerate}[a)]
          \item for a fixed QA coupling, the entanglement measure is a monotonically decreasing function of AB coupling; its values ranges from 1/2 to 0. (Measure decreases means entanglement increases.)

        \item for a fixed AB coupling, the entanglement measure does not always monotonically decrease with increasing QA coupling. It increases first to reach the maximum and then monotonically decreases. This is because part of it results from the behavior of cross-correlations.
        \end{enumerate}

\item Comparison between the two cases:
	\begin{enumerate}[a)]
          \item for QA and QB coupling greater than AB coupling, entanglement for Q vs AB is stronger than entanglement A vs QB.

      \item for AB coupling greater than QA coupling, entanglement for A vs QB is stronger than that for Q vs AB
	\end{enumerate}

\end{itemize}

\subsection{Implications and Applications}

As mentioned in the Introduction, one emergent area of research where results from this analysis can be applied to is macroscopic quantum phenomena.  We identified two key issues to understand, one is  the role of the CoM in the quantum behavior of a many-body system; the second is how the level-of-structure enter in the manifestation of macroscopic quantum phenomena, can be examined using the entanglement structure results obtained here for a three oscillator system with disparate coupling strengths.

Call one of the oscillators Q which is distinguished from the other two A and B by different couplings. Consider the two cases: Case [SYM] when Q is equally coupled to A and B (symmetric configuration) versus Case [ASY] when A is coupled to Q differently from its coupling to B (asymmetric configuration).  In the SYM case we show that as far as quantum entanglement is concerned Q is effectively coupled to the CoM of A and B. This shows that  as far as quantum behavior of this nature is concerned the CoM variable functions similar to its role in classical mechanics.  Now view Q  as the center of mass variable representing a macroscopic object 1, and A, B as representative constituents of a similar macroscopic object 2.  One interesting result we found here of relevance to the level-of-structure aspect of MQP is that (at least for quadratic inter-particle coupling) the entanglement between Q and AB is equivalent to that between Q and the CoM of 2, \textit{independent of the coupling between A and B} provided A and B are interchangeably identically. This is of interest because it explains why as one reports on the quantum dynamics of say a cantilever (e.g., \cite{nem,MQPnem}) one does not need to also provide information of how say, the nucleons in the metal bar, are coupled. Comparison of the results in this case [SYM] with the other case [ASY] is also informative about why levels of structure, or a hierarchical ordering, of matter is a natural way of organization for a better understanding of the manifestation of quantum behavior of macroscopic objects.

The open system of $N$ coupled quantum oscillator system is a generic model for investigating continuous-variables many-body systems under different environmental influence.  Continuing the theme of MQP, one can also  examine the 'size' and the `correlation' aspects in addition to the  entanglement aspects explored here for interacting quantum systems. (These three aspects of MQP are explained in \cite{MQP1,MQP2,MQP3} -- see references therein.) We mention two additional directions currently under study. It can easily be generalized to a finite temperature field whereby one can investigate issues in quantum thermodynamics such as those related to thermal entanglement in \cite{AEPW02,VedralRMP,AndWin,AndWin2,TE2}. One can also partition the $N$ oscillators into two parts: one consisting of $m$ oscillators representing the system and the second part consisting of $n=N-m$ oscillators as the bath. By varying $m$ versus $n$ one can use this model to investigate mesoscopic phenomena \cite{SAI05,TE3,PS01,HH11a,HH11b} addressing new and different facets such as those mentioned above for MQP.

\vskip .5cm

\noindent{\bf Acknowledgment} We are obliged to Prof. S.-Y. Lin for helpful discussions on the covariance matrix structure, Prof.~C.-H. Chou, C. Fleming, and Y. Suba{\c{s}}{\i} for discussions on multi-partite entanglements in relation to macroscopic quantum phenomena. BLH thanks Professors Y. S. Wu and J. Q. You for their kind hospitality at Fudan University in the Spring of 2013 when this work was finished.

\newpage

%\appendix
\section{Appendix: Symmetric Gaussian Systems}\label{S:eriea}
In this appendix, we would like to illustrate two interesting observations in the course of investigating entanglement of the bi-partition system. First, we highlight the role the CoM coordinate in such a system. We then explicitly point out the connection between an orthogonal matrix in the configuration space and the corresponding symplectic matrix in the phase space.

In general, the three-oscillator system, discussed in the previous section, is not a fully symmetric Gaussian system due to disparate inter-oscillator coupling. However, when we make a particular partition such that it is a bipartite $Q$ versus $AB$ system, then we have a $1\times2$ bi-symmetric Gaussian system. We may apply some well-known facts~\cite{SAI05} of the $M\times N$ Gaussian system to this case.

In the case of disparate coupling we considered, if we ignore the influence of the background field, then the Lagrangian corresponding to \eqref{E:jeaiekjna} for $N=3$ can be expressed in terms of the center-of-mass coordinate and the relative coordinate. Following \eqref{E:eriekns}, we write the Lagrangian as
\begin{align}
	 \mathcal{L}&=\sum_{i=1}^{3}\Bigl\{\frac{m}{2}\,\dot{\chi}^{(i)\,2}-\frac{m}{2}\,\omega_{0}^{2}\chi^{(i)\,2}\Bigr\}\notag\\
	 &\qquad\qquad-\frac{m}{2}\,\lambda\bigl(\chi^{(1)}-\chi^{(2)}\bigr)^{2}-\frac{m}{2}\,\lambda\bigl(\chi^{(1)}-\chi^{(3)}\bigr)^{2}-\frac{m}{2}\,\sigma\bigl(\chi^{(2)}-\chi^{(3)}\bigr)^{2}\notag\\
	 &=\Bigl\{\frac{m}{2}\,\dot{\chi}^{(1)\,2}-\frac{m}{2}\,\bigl(\omega_{0}^{2}+2\lambda\bigr)\,\chi^{(1)\,2}\Bigr\}+m\lambda\,\chi^{(1)}\bigl(\chi^{(2)}+\chi^{(3)}\bigr)\notag\\
	 &\qquad\qquad+\Bigl\{\frac{m}{4}\,\bigl(\dot{\chi}^{(2)}+\dot{\chi}^{(3)}\bigr)^{2}-\frac{m}{4}\,\bigl(\omega_{0}^{2}+\lambda\bigr)\bigl(\chi^{(2)}+\chi^{(3)}\bigr)^{2}\Bigr\}\notag\\
	 &\qquad\qquad+\Bigl\{\frac{m}{4}\,\bigl(\dot{\chi}^{(2)}-\dot{\chi}^{(3)}\bigr)^{2}-\frac{m}{4}\,\bigl(\omega_{0}^{2}+\lambda+2\sigma\bigr)\bigl(\chi^{(3)}-\chi^{(3)}\bigr)^{2}\Bigr\}\,.\label{E:nerkje}
\end{align}
From this construction we see that the degree of freedom $\chi^{(2)}-\chi^{(3)}$ is decoupled from the remaining degrees of freedom. In addition, only the center-of-mass coordinate $(\chi^{(2)}+\chi^{(3)})/2$ for the subsystem 2 and 3 couple with the coordinate $\chi_{1}$ of the system 1. Thus during the course of evolution, the degree of freedom $\chi^{(2)}-\chi^{(3)}$ is always separate from the others, and may not be involved in creation of entanglement of the whole system. For example, in the case of $Q$ versus $AB$, the entanglement is obtained by partial transpose the density matrix associated with the subsystem 1, that is $Q$, so the degree of freedom $\chi^{(2)}-\chi^{(3)}$ remains uninvolved. Hence the entanglement for $Q$ versus $AB$ is completely determined and quantified by the degrees of freedom $\chi^{(1)}$ and $(\chi^{(2)}+\chi^{(3)})/2$. From \eqref{E:nerkje}, we can see that the entanglement measure should not depend on the coupling constant $\sigma$ between subsystem 2 and 3 because it does not appear in the first line of eq.~\eqref{E:nerkje}.

On the other hand, when we make a partition like $A$ versus $QB$, the coordinate transformation of $\chi^{(1)}$, $\chi^{(2)}$ and $\chi^{(3)}$ into $\chi^{(1)}$, $(\chi^{(2)}+\chi^{(3)})/2$ and $\chi^{(2)}-\chi^{(3)}$ does not have any advantageous convenience because it does not conform to any symmetry this particular partition renders. In addition, when we make partial transpose on either subsystem 2 (that is $A$), or subsystem 3 (that is $B$), both degrees of freedom $(\chi^{(2)}+\chi^{(3)})/2$, $\chi^{(2)}-\chi^{(3)}$ will be involved. Thus in this case, the entanglement measure for $A$ versus $QB$ will inevitably depend on $\sigma$.

We would like to generalize the system introduced in \eqref{E:nerkje} to a $1\times(N-1)$ fully symmetric system. Let the corresponding Lagrangian be given by
\begin{align}
	 \mathcal{L}&=\Bigl\{\frac{m}{2}\,\dot{\chi}^{(1)\,2}-\frac{m}{2}\,\omega_{0}^{2}\,\chi^{(1)\,2}\Bigr\}-\sum_{i=2}^{N}\frac{m}{2}\,\lambda\bigl(\chi^{(1)}-\chi^{(i)}\bigr)^{2}\label{E:lwnrb}\\
	 &\qquad\qquad\qquad\qquad+\sum_{i=2}^{N}\Bigl\{\frac{m}{2}\,\dot{\chi}^{(i)\,2}-\frac{m}{2}\,\omega_{0}^{2}\,\chi^{(i)\,2}-\sum_{j>i}^{N}\frac{m}{2}\,\sigma\bigl(\chi^{(i)}-\chi^{(j)}\bigr)^{2}\Bigr\}\,.\notag
\end{align}
It describe a specially partitioned $N$-component system which consists of two groups, subsystem 1 and the rest of the components. The subsystem 1 has the same coupling strength $\lambda$ with the the rest of the subsystems, meanwhile the remaining subsystems interact with one another with the same coupling constant $\sigma$. Apparently this is a fully symmetric $1\times(N-1)$ system because it does not change when we permute $\chi_{i}$ for $i=2,\ldots,N$. It is a generalization of $Q$ versus $AB$ configuration to $Q$ versus $A_{2}A_{3}\cdots A_{N}$.

Recall that we can always find the normal modes for a multi-component system described by the Lagrangian \eqref{E:lwnrb}. We first rewrite the Lagrangian \eqref{E:lwnrb} as
\begin{align}
	 \mathcal{L}&=\Bigl\{\frac{m}{2}\,\dot{\chi}^{(1)\,2}-\frac{m}{2}\Bigl[\omega_{0}^{2}+(N-1)\lambda\Bigr]\chi^{(1)\,2}\Bigr\}+\frac{m}{2}\,\lambda\,\chi^{(1)}\sum_{i=2}^{N}\chi^{(i)}\notag\\
	 &\qquad\qquad\qquad+\sum_{i=2}^{N}\Bigl\{\frac{m}{2}\,\dot{\chi}^{(i)\,2}-\frac{m}{2}\Bigl[\omega_{0}^{2}+\lambda+(N-2)\sigma\Bigr]\chi^{(i)\,2}\Bigr\}+\sum_{j\neq i=2}^{N}\frac{m}{2}\,\sigma\,\chi^{(i)}\chi^{(j)}\notag\\
	 &=\Bigl\{\frac{m}{2}\,\dot{\chi}^{(1)\,2}-\frac{m}{2}\Bigl[\omega_{0}^{2}+(N-1)\lambda\Bigr]\chi^{(1)\,2}\Bigr\}\notag\\
	 &\qquad\qquad\qquad+\frac{m}{2}\,\lambda\,\chi^{(1)}\sum_{i=2}^{N}\chi^{(i)}+\frac{m}{2}\mathbf{y}^{T}\cdot\mathbf{y}-\frac{m}{2}\,\mathbf{y}^{T}\cdot\pmb{\Omega}^{2}\cdot\mathbf{y}\,,\label{E:yuwkwz}
\end{align}
where $\mathbf{y}=(\chi^{(2)},\ldots,\chi^{(N)})^{T}$, and
\begin{equation}
	\pmb{\Omega}^{2}=\begin{pmatrix}\omega_{0}^{2}+\lambda+(N-2)\sigma &-\sigma &\cdots &-\sigma\\
	-\sigma &\omega_{0}^{2}+\lambda+(N-2)\sigma &\cdots &-\sigma\\
	\vdots  &\vdots &\ddots &\vdots\\
	-\sigma &-\sigma &\cdots &\omega_{0}^{2}+\lambda+(N-2)\sigma\end{pmatrix}\,.
\end{equation}
This $(N-1)\times(N-1)$ interaction matrix has an interesting feature that when we sum elements along either one of the rows or columns, we obtain the same result. This feature, as discussed before \eqref{E:oeruohnw}, implies that one of the normal modes must be of the form
\begin{equation}
	\xi^{(2)}=\frac{1}{\sqrt{N-1}}\sum_{i=2}^{N}\chi^{(i)}\,,
\end{equation}
which apparently is the center of mass coordinate for the original variables $\chi^{(2)}$,..., $\chi^{(N)}$. If we let $\xi^{(j)}$, for $j=2,\ldots,N$ be the normal mode coordinates, then the Lagrangian \eqref{E:yuwkwz} becomes
\begin{align}
	 \mathcal{L}&=\Bigl\{\frac{m}{2}\,\dot{\chi}^{(1)\,2}-\frac{m}{2}\Bigl[\omega_{0}^{2}+(N-1)\lambda\Bigr]\chi^{(1)\,2}\Bigr\}+\frac{m}{2}\,\sqrt{N-1}\lambda\,\chi^{(1)}\xi^{(2)}\notag\\
	 &\qquad\qquad\qquad+\frac{m}{2}\,\dot{\xi}^{(2)\,2}-\frac{m}{2}\,\omega_{2}^{2}\,\xi^{(2)\,2}+\sum_{j=3}^{N}\Bigl\{\frac{m}{2}\,\dot{\xi}^{(j)\,2}-\frac{m}{2}\,\omega_{j}^{2}\,\xi^{(j)\,2}\Bigr\}\,,
\end{align}
where $\omega_{j}$ is the eigenfrequency associated with the normal mode $\xi^{(j)}$, or in other words
\begin{equation}
	\pmb{\Lambda}^{2}=\begin{pmatrix}\omega_{2}^{2} &0 &0 &\cdots &0\\[4pt]0 &\omega_{3}^{2} &0 &\cdots &0\\[4pt]
	0 &0 &\ddots & &\vdots\\[4pt]
	\vdots &\vdots & &\ddots &\vdots\\[4pt]
	0 &0 &\cdots &0 &\omega_{N}^{2}\end{pmatrix}\,.
\end{equation}
For the later reference, $\omega_{2}^{2}$ is given by $\omega_{2}^{2}=\sum_{i}[\pmb{\Omega}^{2}]_{ij}=\omega_{0}^{2}+\lambda$ and we have $(N-2)$ degenerate eigenfrequencies $\omega_{3}^{2}=\cdots=\omega_{N}^{2}=\omega_{0}^{2}+\lambda+(N-1)\sigma$. Therefore following the previous arguments, if we compute entanglement for the partition $Q$ versus $A_{2}A_{3}\cdots A_{N}$, then only these two degrees of freedom are involved in creation of entanglement and the corresponding entanglement measure will not depend on $\sigma$. In fact, the remaining $(N-2)$ normal modes $\xi^{(k)}$, for $k=3,\ldots,N$, remain decoupled from one another and evolve with time by themselves. Essentially only $\chi^{(1)}$ and $\xi^{(2)}$ interact together and evolve into an entangled state. Furthermore their evolution does not depend on the details of evolution of the degrees of freedom $\xi^{(k)}$. That is, $\xi^{(k)}$ will not contribute to entanglement between $\chi^{(1)}$ and $\xi^{(2)}$. The covariance matrix of the whole system in terms of $\chi^{(1)}$ and the normal modes $\xi^{(j)}$, $j=2,3,\ldots,N$, will look like
\begin{equation}\label{E:nkuryw}
	\pmb{\sigma}'=\begin{pmatrix}\pmb{\alpha}_{1} &\pmb{\gamma} &\mathbf{0} &\cdots &\mathbf{0}\\[4pt]
	\pmb{\gamma}^{T} &\pmb{\beta}_{2} &\mathbf{0} &\cdots &\mathbf{0}\\[4pt]
	\mathbf{0} &\mathbf{0} &\pmb{\beta}_{3} & &\mathbf{0}\\[4pt]
	\vdots &\vdots & &\ddots &\vdots\\[4pt]
	\mathbf{0} &\mathbf{0} &\mathbf{0} &\cdots &\pmb{\beta}_{N}
	\end{pmatrix}
\end{equation}
Not that this is not the Williamson form of the covariance matrix. The $2\times2$ matrix $\pmb{\gamma}$ account for cross-correlation between $\chi^{(1)}$ and $\xi^{(2)}$. The covariance matrix \eqref{E:nkuryw} happens to take the ``localized'' form as demonstrated in~\cite{SAI05}, that is, the center-of-mass coordinate $\xi^{(2)}$ is the representative mode of the $N-1$ fully symmetric subsystems, responsible for setting up correlation with the other party, represented by $\chi^{(1)}$.

Then how does this observation go along with the procedures of diagonalizing the covariance matrix for a $M\times N$ bi-symmetric system, discussed in~\cite{SAI05}? We again use the example of the $1\times2$ system with disparate coupling for illustration. We will closely follow the procedures outlined in~\cite{SAI05}. Generically speaking the covariance matrix for our tripartite system \eqref{E:eriekns} takes the form
\begin{equation}\label{E:gfnekr}
	\pmb{\sigma}_{1\times2}=\begin{pmatrix}a &0 &b &0 &b &0\\[4pt]0 &e &0 &f &0 &f\\[4pt]b &0 &c &0 &d &0\\[4pt]0 &f &0 &g &0 &h\\[4pt]b &0 &d &0 &c &0\\[4pt]0 &f &0 &h &0 &g\end{pmatrix}\,,
\end{equation}
where
\begin{align*}
	 a&=\frac{1}{2}\,\langle\,\{\chi_{b}^{(1)},\chi_{b}^{(1)}\}\,\rangle\,,&b&=\frac{1}{2}\,\langle\,\{\chi_{b}^{(1)},\chi_{b}^{(2)}\}\,\rangle=\langle\,\{\chi_{b}^{(1)},\chi_{b}^{(3)}\}\,\rangle\,,\\
	 c&=\frac{1}{2}\,\langle\,\{\chi_{b}^{(2)},\chi_{b}^{(2)}\}\,\rangle=\langle\,\{\chi_{b}^{(3)},\chi_{b}^{(3)}\}\,\rangle\,,&d&=\frac{1}{2}\,\langle\,\{\chi_{b}^{(2)},\chi_{b}^{(3)}\}\,\rangle\,,\\
	 e&=\frac{1}{2}\,\langle\,\{p_{b}^{(1)},p_{b}^{(1)}\}\,\rangle\,,&f&=\frac{1}{2}\,\langle\,\{p_{b}^{(1)},p_{b}^{(2)}\}\,\rangle=\langle\,\{p_{b}^{(1)},p_{b}^{(3)}\}\,\rangle\,,\\
	 g&=\frac{1}{2}\,\langle\,\{p_{b}^{(2)},p_{b}^{(2)}\}\,\rangle=\langle\,\{p_{b}^{(3)},p_{b}^{(3)}\}\,\rangle\,,&h&=\frac{1}{2}\,\langle\,\{p_{b}^{(2)},p_{b}^{(3)}\}\,\rangle\,,
\end{align*}
as shown in \eqref{E:oeowjx}--\eqref{E:neowjx} and \eqref{E:hyeiw}. Next we would like to write the covariance matrix by a series symplectic transformations into the standard form
\begin{equation}
	\pmb{\sigma}_{1\times2}^{(1)}=\begin{pmatrix}\pmb{\alpha} &\pmb{\gamma} &\pmb{\gamma}\\\pmb{\gamma} &\pmb{\beta} &\pmb{\epsilon}\\\pmb{\gamma} &\pmb{\epsilon} &\pmb{\beta}\end{pmatrix}
\end{equation}
where $\pmb{\alpha}$, $\pmb{\beta}$, $\pmb{\epsilon}$, $\pmb{\gamma}$ are $2\times2$ block matrices, and take the form
\begin{align*}
	\pmb{\alpha}&=\begin{pmatrix}\alpha &0\\[4pt]0 &\alpha\end{pmatrix}\,,&\pmb{\beta}&=\begin{pmatrix}\beta &0\\[4pt]0 &\beta\end{pmatrix}\,, &\pmb{\epsilon}=\begin{pmatrix}\epsilon_{1} &0\\[4pt]0 &\epsilon_{2}\end{pmatrix}\,.
\end{align*}
Since for our simple example, \eqref{E:gfnekr} is already closed to standard form, what we will do is to apply a symplectic transformation $S_{1}$, which is equivalent to rescaling or change of the units, upon $\pmb{\sigma}$ such that the $2\times2$ block matrices along the diagonal of $\pmb{\sigma}$ are proportional the identity matrix. For lower dimensional matrix, this transformation matrix can be found without much effort. Let us take the upper left block of $\pmb{\sigma}_{1\times2}$ for example. Let the $2\times2$ matrix $\mathbf{A}$ denote
\begin{equation*}
	\mathbf{A}=\begin{pmatrix}a &0\\0 &e\end{pmatrix}\,,
\end{equation*}
and then the corresponding $2\times2$ transformation matrix $\mathbf{S}$
\begin{equation}
	\mathbf{S}=\begin{pmatrix}\mu &\lambda\\\kappa &\nu\end{pmatrix}
\end{equation}
can be obtained by requiring that
\begin{equation}
	\mathbf{S}\cdot\pmb{\omega}\cdot\mathbf{S}^{T}=\pmb{\omega}\,,\qquad\qquad\qquad\pmb{\omega}=\begin{pmatrix}0 &1\\-1 &0\end{pmatrix}\,,
\end{equation}
and that
\begin{equation}
	\mathbf{S}\cdot\mathbf{A}\cdot\mathbf{S}^{T}
\end{equation}
should be proportional to a $2\times2$ identity matrix. We then find
\begin{equation}
  	\mathbf{S}=\begin{pmatrix}\sqrt[4]{\dfrac{e}{a}} &0\\[10pt]0 &\sqrt[4]{\dfrac{a}{e}}\end{pmatrix}\,.
\end{equation}
We can do the same thing to the remain $2\times2$ block matrices along the diagonal. Thus the symplectic matrix $S_{1}$, which we use to diagonalize $\pmb{\sigma}_{1\times2}$, will take the form
\begin{equation}
	S_{1}=\begin{pmatrix}
	\sqrt[4]{\dfrac{e}{a}} &0 &0 &0 &0 &0\\[10pt]
	0 &\sqrt[4]{\dfrac{a}{e}} &0 &0 &0 &0\\[10pt]
	0 &0 &\sqrt[4]{\dfrac{g}{c}} &0 &0 &0\\[10pt]
	0 &0 &0 &\sqrt[4]{\dfrac{c}{g}} &0 &0\\[10pt]
	0 &0 &0 &0 &\sqrt[4]{\dfrac{g}{c}} &0\\[10pt]
	0 &0 &0 &0 &0 &\sqrt[4]{\dfrac{c}{g}}
	\end{pmatrix}\,,
\end{equation}
and then
\begin{equation}
	\pmb{\sigma}_{1\times2}^{(1)}=S_{1}^{\vphantom{T}}\cdot\pmb{\sigma}_{1\times2}\cdot S_{1}^{T}=\begin{pmatrix}
	\sqrt{ae} &0 &b\sqrt[4]{\dfrac{eg}{ac}} &0 &b\sqrt[4]{\dfrac{eg}{ac}} &0\\[10pt]
	0 &\sqrt{ae} &0 &f\sqrt[4]{\dfrac{ac}{eg}} &0 &f\sqrt[4]{\dfrac{ac}{eg}}\\[10pt]
	b\sqrt[4]{\dfrac{eg}{ac}} &0 &\sqrt{cg} &0 &\sqrt{\dfrac{g}{c}}\,d &0\\[10pt]
	0 &f\sqrt[4]{\dfrac{ac}{eg}} &0 &\sqrt{cg} &0 &\sqrt{\dfrac{c}{g}}\,h\\[10pt]
	b\sqrt[4]{\dfrac{eg}{ac}} &0 &\sqrt{\dfrac{g}{c}}\,d &0 &\sqrt{cg} &0\\[10pt]
	0 &f\sqrt[4]{\dfrac{ac}{eg}} &0 &\sqrt{\dfrac{c}{g}}\,h &0 &\sqrt{cg}\end{pmatrix}\,.
\end{equation}
Since the upper left $2\times2$ block matrix of $\pmb{\sigma}_{1\times2}^{(1)}$ is already diagonalized and is in its Williamson form, next we will diagonalize the lower right $4\times4$ block matrix
\begin{equation}
	\mathbf{F}=\begin{pmatrix}\sqrt{cg} &0 &\sqrt{\dfrac{g}{c}}\,d &0\\[10pt]0 &\sqrt{cg} &0 &\sqrt{\dfrac{c}{g}}\,h\\[10pt]\sqrt{\dfrac{g}{c}}\,d &0 &\sqrt{cg} &0\\[10pt]0 &\sqrt{\dfrac{c}{g}}\,h &0 &\sqrt{cg}\end{pmatrix}
\end{equation}
into the Williamson normal form.

We may follow the procedures outlined in~\cite{SAI05} to find the Williamson form; however it becomes a little impractical in reality due to the ambiguity that we can always introduce an additional orthogonal matrix when we construct a symplectic matrix that transforms a given symmetric matrix into its Williamson form. Instead, we make an observation that a symplectic transformation in the phase space has its corresponding rotation transformation in the configuration space. In addition, in quantum mechanics the canonical momentum operator is defined in such a way that it transforms in the same manners as the position operator does. Therefore, to avoid cluttering of the square root notations, we write the matrix $\mathbf{F}$ as
\begin{equation}\label{E:bfhwkw}
	\mathbf{F}=\begin{pmatrix}a_{1} &0 &b_{1} &0\\[4pt]0 &a_{1} &0 &c_{1}\\[4pt]b_{1} &0 &a_{1} &0\\[4pt]0 &c_{1} &0 &a_{1}\end{pmatrix}
\end{equation}
and suppose that the position column vector $(\chi_{2},\chi_{3})^{T}$ is rotated by an orthogonal matrix
\begin{equation}
	\begin{pmatrix}\cos\theta &\sin\theta\\[10pt]-\sin\theta &\cos\theta\end{pmatrix}\,,	
\end{equation}
so that the corresponding symplectic transformation matrix
\begin{equation}
	\mathbf{R}=\begin{pmatrix}\cos\theta &0 &\sin\theta &0\\[4pt]0 &\cos\theta &0 &\sin\theta\\[4pt]-\sin\theta &0 &\cos\theta &0\\[4pt]0 &-\sin\theta &0 &\cos\theta\end{pmatrix}
\end{equation}
makes the matrix $\mathbf{F}$ into the diagonal form. Of course this does not guarantee that the resulting diagonal matrix is in the Williamson form. The requirement of
\begin{equation*}
	\mathbf{R}\cdot\mathbf{F}\cdot\mathbf{R}^{T}
\end{equation*}
being a diagonal matrix gives $\theta=\pm\pi/4$, and the resulting matrix is
\begin{equation}
	\mathbf{F}_{1}=\mathbf{R}\cdot\mathbf{F}\cdot\mathbf{R}^{T}=\begin{pmatrix}a_{1}+b_{1} &0 &0 &0\\[4pt]0 &a_{1}+c_{1} &0 &0\\[4pt]0 &0 &a_{1}-b_{1} &0\\[4pt]0 &0 &0 &a_{1}-c_{1}\end{pmatrix}\,,
\end{equation}
if we choose $\theta=\pi/4$. To make this into the Williamson form, we again apply another transformation $\mathbf{S}'_{1}$ to rescale the diagonal elements,
\begin{equation}
	\mathbf{T}_{1}=\begin{pmatrix}\sqrt[4]{\dfrac{a_{1}+c_{1}}{a_{1}+b_{1}}} &0 &0 &0\\[8pt]0 &\sqrt[4]{\dfrac{a_{1}+b_{1}}{a_{1}+c_{1}}} &0 &0\\[8pt]0 &0 &\sqrt[4]{\dfrac{a_{1}-c_{1}}{a_{1}-b_{1}}} &0\\[8pt]0 &0 &0 &\sqrt[4]{\dfrac{a_{1}-b_{1}}{a_{1}-c_{1}}}\end{pmatrix}
\end{equation}
so that
\begin{equation}
	 \mathbf{T}_{1}^{\vphantom{T}}\cdot\mathbf{F}_{1}\cdot\mathbf{T}_{1}^{T}=\mathbf{T}_{1}^{\vphantom{T}}\cdot\mathbf{R}\cdot\mathbf{F}\cdot\mathbf{R}^{T}\cdot\mathbf{T}_{1}^{T}=\begin{pmatrix}\eta_{3} &0 &0 &0\\[8pt]0 &\eta_{3} &0 &0\\[8pt]0 &0 &\eta_{2} &0\\[8pt]0 &0 &0 &\eta_{2}\end{pmatrix}\,.
\end{equation}
where $\eta_{2}=\sqrt{(a_{1}-b_{1})(a_{1}-c_{1})}$ and $\eta_{3}=\sqrt{(a_{1}+b_{1})(a_{1}-c_{1})}$. Now we have arrived at the Williamson form of the matrix $\mathbf{F}$, \eqref{E:bfhwkw}. The diagonal elements are its symplectic eigenvalues, and can be verified by solving the secular equation of the corresponding symplectic eigenvalue problem
\begin{equation}
	\det\bigl(\mathbf{F}-i\,\eta\,\Omega_{2}\bigr)=0\,,\qquad\qquad\qquad\Omega_{2}=\bigoplus_{k=1}^{2}\pmb{\omega}\,.
\end{equation}
Now putting these results together, we have the symplectic matrix $S_{2}$
\begin{equation}\label{E:fhekssa}
	S_{2}=\begin{pmatrix}1 &0 &0 &0 &0 &0\\0 &1 &0 &0 &0 &0\\0 &0 &\dfrac{1}{\sqrt{2}}\sqrt[4]{\dfrac{a_{1}+c_{1}}{a_{1}+b_{1}}} &0 &\dfrac{1}{\sqrt{2}}\sqrt[4]{\dfrac{a_{1}+c_{1}}{a_{1}+b_{1}}} &0\\[8pt]0 &0 &0 &\dfrac{1}{\sqrt{2}}\sqrt[4]{\dfrac{a_{1}+b_{1}}{a_{1}+c_{1}}} &0 &\dfrac{1}{\sqrt{2}}\sqrt[4]{\dfrac{a_{1}+b_{1}}{a_{1}+c_{1}}}\\[8pt] 0 &0 &-\dfrac{1}{\sqrt{2}}\sqrt[4]{\dfrac{a_{1}-c_{1}}{a_{1}-b_{1}}} &0 &\dfrac{1}{\sqrt{2}}\sqrt[4]{\dfrac{a_{1}-c_{1}}{a_{1}-b_{1}}} &0\\[8pt]0 &0 &0 &-\dfrac{1}{\sqrt{2}}\sqrt[4]{\dfrac{a_{1}-b_{1}}{a_{1}-c_{1}}} &0 &\dfrac{1}{\sqrt{2}}\sqrt[4]{\dfrac{a_{1}-b_{1}}{a_{1}-c_{1}}}\end{pmatrix}\,,
\end{equation}
with
\begin{align*}
	a_{1}&=\sqrt{cg}\,,& b_{1}&=\sqrt{\dfrac{g}{c}}\,d\,, &c_{1}&=\sqrt{\dfrac{c}{g}}\,h\,.
\end{align*}
It will transform the lower right $4\times4$ block matrix of $\pmb{\sigma}_{1\times2}^{(1)}$ into its Williamson form
\begin{equation}\label{E:erkebwq}
	\pmb{\sigma}_{1\times2}^{(2)}=S_{2}^{\vphantom{T}}\cdot\pmb{\sigma}_{1\times2}^{(1)}\cdot S_{2}^{T}=\begin{pmatrix}\eta_{\circ} &0 &\xi &0 &0 &0\\[8pt]0 &\eta_{\circ} &0 &\zeta &0 &0\\[8pt]\xi &0 &\eta_{-} &0 &0 &0\\[8pt]0 &\zeta &0 &\eta_{-} &0 &0\\[8pt]0 &0 &0 &0 &\eta_{+} &0\\[8pt]0 &0 &0 &0 &0 &\eta_{+}\end{pmatrix}\,.
\end{equation}
with
\begin{align*}
	\eta_{\circ}&=\sqrt{ae}\,, &\eta_{-}&=\sqrt{(c-d)(g-h)}\,, &\eta_{+}&=\sqrt{(c+d)(g+h)}\,, \\\xi&=\sqrt{2}b\,\sqrt[4]{\dfrac{e(g+h)}{a(c+d)}}\,, &\zeta&=\sqrt{2}f\,\sqrt[4]{\dfrac{a(c+d)}{e(g+h)}}\,.
\end{align*}
Note that $\eta_{\circ}$, $\eta_{\pm}$ shown above are not the symplectic eigenvalues of the $6\times6$ matrix $\pmb{\sigma}_{1\times2}$. Instead $\eta_{\circ}$ is the symplectic eigenvalue of the upper left $2\times2$ matrix but $\eta_{\pm}$ are the symplectic eigenvalues of the lower right $4\times4$ matrix. The true symplectic eigenvalues of the $6\times6$ matrix are in fact given by
\begin{align}
	\eta_{1}&=\sqrt{(c-d)(g-h)}\,,\\
	\eta_{2,\,3}&=\Biggl\{\frac{ae+4bf+(c+d)(g+h)}{2}\Biggr.\\
	 &\quad\pm\Biggl.\sqrt{\left[\frac{ae+4bf+(c+d)(g+h)}{2}\right]^{2}-\Bigl[2b^{2}-a(c+d)\Bigr]\Bigl[2f^{2}-e(g+h)\Bigr]}\Biggr\}^{1/2}\,,\notag
\end{align}
so only $\eta_{-}$ coincides with one of the symplectic eigenvalues of the covariance matrix. The result in \eqref{E:erkebwq} shows that the cross-correlation between the subsystem 1 ($Q$) and the union of the subsystems 2 \& 3 ($AB$) is linked up by one representative degree of freedom from each partition. This correlation is described by the block matrix
\begin{equation}
	\pmb{\gamma}''=\begin{pmatrix}\xi &0\\0 &\zeta\end{pmatrix}\,.
\end{equation}
It will be interesting to see what degrees of freedom are involved in cross-correlation between two partitions. To do so, we first combine $S_{2}$ and $S_{1}$ together
\begin{equation*}
	S=S_{2}\cdot S_{1}=\begin{pmatrix}
		\sqrt[4]{\dfrac{e}{a}} &0 &0 &0 &0 &0\\[10pt]
		0 &\sqrt[4]{\dfrac{a}{e}} &0 &0 &0 &0\\[10pt]
		0 &0 &\dfrac{1}{\sqrt{2}}\sqrt[4]{\dfrac{g+h}{c+d}}  &0 &\dfrac{1}{\sqrt{2}}\sqrt[4]{\dfrac{g+h}{c+d}}  &0\\[10pt]
		0 &0 &0 &\dfrac{1}{\sqrt{2}}\sqrt[4]{\dfrac{c+d}{g+h}}  &0 &\dfrac{1}{\sqrt{2}}\sqrt[4]{\dfrac{c+d}{g+h}}\\[10pt]
		0 &0 &-\dfrac{1}{\sqrt{2}}\sqrt[4]{\dfrac{g-h}{c-d}}  &0 &\dfrac{1}{\sqrt{2}}\sqrt[4]{\dfrac{g-h}{c-d}}  &0\\[10pt]
		0 &0 &0 &-\dfrac{1}{\sqrt{2}}\sqrt[4]{\dfrac{c-d}{g-h}}  &0 &\dfrac{1}{\sqrt{2}}\sqrt[4]{\dfrac{c-d}{g-h}}\end{pmatrix}
\end{equation*}
and apply the transformation matrix $S$ to the column vector $(\chi^{(1)},p^{(1)},\chi^{(2)},p^{(2)},\chi^{(3)},p^{(3)})^{T}$, which describes the state of the tripartite system. It gives
\begin{align}
	&\quad\begin{pmatrix}
		\sqrt[4]{\dfrac{e}{a}} &0 &0 &0 &0 &0\\[10pt]
		0 &\sqrt[4]{\dfrac{a}{e}} &0 &0 &0 &0\\[10pt]
		0 &0 &\dfrac{1}{\sqrt{2}}\sqrt[4]{\dfrac{g+h}{c+d}}  &0 &\dfrac{1}{\sqrt{2}}\sqrt[4]{\dfrac{g+h}{c+d}}  &0\\[10pt]
		0 &0 &0 &\dfrac{1}{\sqrt{2}}\sqrt[4]{\dfrac{c+d}{g+h}}  &0 &\dfrac{1}{\sqrt{2}}\sqrt[4]{\dfrac{c+d}{g+h}}\\[10pt]
		0 &0 &-\dfrac{1}{\sqrt{2}}\sqrt[4]{\dfrac{g-h}{c-d}}  &0 &\dfrac{1}{\sqrt{2}}\sqrt[4]{\dfrac{g-h}{c-d}}  &0\\[10pt]
		0 &0 &0 &-\dfrac{1}{\sqrt{2}}\sqrt[4]{\dfrac{c-d}{g-h}}  &0 &\dfrac{1}{\sqrt{2}}\sqrt[4]{\dfrac{c-d}{g-h}}\end{pmatrix}\cdot\begin{pmatrix}\chi^{(1)}\vphantom{\bigg|}\\[12pt]p^{(1)}\vphantom{\bigg|}\\[12pt]\chi^{(2)}\vphantom{\bigg|}\\[12pt]p^{(2)}\vphantom{\bigg|}\\[12pt]\chi^{(3)}\vphantom{\bigg|}\\[12pt]p^{(3)}\vphantom{\bigg|}\end{pmatrix}\notag\\
		 &=\begin{pmatrix}\sqrt[4]{\dfrac{e}{a}}\;\chi^{(1)}\vphantom{\bigg|}\\[12pt]\sqrt[4]{\dfrac{a}{e}}\;p^{(1)}\vphantom{\bigg|}\\[12pt]\sqrt[4]{\dfrac{g+h}{c+d}}\;\dfrac{\chi^{(2)}+\chi^{(3)}}{\sqrt{2}}\vphantom{\bigg|}\\[12pt]\sqrt[4]{\dfrac{c+d}{g+h}}\;\dfrac{p^{(2)}+p^{(3)}}{\sqrt{2}}\vphantom{\bigg|}\\[12pt]-\sqrt[4]{\dfrac{g-h}{c-d}}\;\dfrac{\chi^{(2)}-\chi^{(3)}}{\sqrt{2}}\vphantom{\bigg|}\\[12pt]-\sqrt[4]{\dfrac{c-d}{g-h}}\;\dfrac{p^{(2)}-p^{(3)}}{\sqrt{2}}\vphantom{\bigg|}\end{pmatrix}\,.\label{E:ernkwiq}
\end{align}
Again from this and \eqref{E:erkebwq}, we explicitly show that the cross-correlation between $Q$ and $AB$ are set up by the degrees of freedom $\chi^{(1)}$ and $[\chi^{(2)}+\chi^{(3)}]/2$. Here we emphasize that this decomposition is suitable only for entanglement of $Q$ with $AB$.

We may want to perform further transformations to get rid of the factors before $\chi^{(1)}$, $p^{(1)}$, $\chi^{(2)}+\chi^{(3)}$, and etc. in \eqref{E:ernkwiq}. However once we do so the resulting covariance matrix no longer takes the Williamson form. In addition, we may observe that those factors play the role of rescaling or changes of the unit in such a way that all elements in the column vector have the same footing or unit.

A comment is ready. We may use the idea of normal modes to help find the Williamson form of a given covariance matrix. Suppose the $N\times N$ matrix $\mathbf{U}$ is composed of the orthonormal eigenvectors of the interaction matrix $\pmb{\Omega}^{2}$, as shown in \eqref{E:eriekns}. We know that the matrix $\mathbf{U}$ will diagonalize $\pmb{\Omega}^{2}$ into $\Lambda^{2}=\mathbf{U}^{T}\cdot\pmb{\Omega}^{2}\cdot\mathbf{U}$ and the normal mode coordinates will be given by $\mathbf{U}^{T}\cdot\mathbf{y}$, where $\mathbf{y}=(\chi^{(1)},\chi^{(2)},\ldots,\chi^{(N)})^{T}$. We may construct a $2N\times 2N$ symplectic transformation matrix $\mathbf{S}$ based on $\mathbf{U}$ such that
\begin{equation}\label{E:erbwks}
	\mathbf{S}=\begin{pmatrix}
		U_{11} &0 &U_{12} &0 &\cdots &\cdots &U_{1N} &0\\
		0 &U_{11} &0 &U_{12} &\cdots &\cdots &0 &U_{1N}\\
		U_{21} &0 &U_{22} &0 &\cdots &\cdots &U_{2N} &0\\
		0 &U_{21} &0 &U_{22} &\cdots &\cdots &0 &U_{2N}\\
		\vdots & & & &\ddots & & &\vdots\\
		\vdots & & & & &\ddots & &\vdots\\
		U_{N1} &0 &U_{N2} &0 &\cdots &\cdots &U_{NN} &0\\
		0 &U_{N1} &0 &U_{N2} &\cdots &\cdots &0 &U_{NN}\end{pmatrix}\,.
\end{equation}
This transformation matrix will put the original covariance matrix into an diagonal form
\begin{equation}
	\widetilde{\pmb{\sigma}}=\mathbf{S}^{T}\cdot\pmb{\sigma}\cdot\mathbf{S}\,.
\end{equation}
Recall that there is not correlation between the position and the canonical momentum of a free oscillator. The diagonal elements of $\widetilde{\pmb{\sigma}}$ consists of the position and the momentum uncertainty of the normal modes. With appropriate scaling or change of the units, it becomes the Williamson form.

However, this form is not very useful for computing the entanglement measure associated with the partial transpose, because even though the Williamson form $\widetilde{\pmb{\sigma}}$ is related to its original covariance matrix $\pmb{\sigma}$ by the symplectic transformation, their partial-transposed counterparts $\widetilde{\pmb{\sigma}}^{pt}$, $\pmb{\sigma}^{pt}$ are not. In other words, the partial transpose of the Williamson form of the covariance matrix $\pmb{\sigma}$ is not necessarily the same as the Williamson form of the partial transpose of the same covariance matrix. Thus the entanglement endowed in the original covariance can not be recovered from the partial transpose of its Williamson form. For example, let the matrix $\mathbf{J}$ account for flipping the sign of $p^{(1)}$, that is
\begin{equation}
	\mathbf{J}=\begin{pmatrix}1 &0 &0 &0 &0 &0\\
	0 &-1 &0 &0 &0 &0\\
	0 &0 &1 &0 &0 &0\\
	0 &0 &0 &1 &0 &0\\
	0 &0 &0 &0 &1 &0\\
	0 &0 &0 &0 &0 &1\end{pmatrix}\,.
\end{equation}
We note that the transformation represented by $\mathbf{J}$ is not a symplectic transformation because
\begin{equation}
	\mathbf{J}\cdot\Omega\cdot\mathbf{J}\neq\Omega\,.
\end{equation}
This implies that the covariance matrix corresponding to the partial transpose with respect to subsystem 1 is given by
\begin{equation}
	 \pmb{\sigma}^{pt}=\mathbf{J}\cdot\pmb{\sigma}\cdot\mathbf{J}^{T}=\mathbf{J}\cdot\mathbf{S}\cdot\widetilde{\pmb{\sigma}}\cdot\mathbf{S}^{T}\cdot\mathbf{J}^{T}=\mathbf{J}\cdot\mathbf{S}\cdot\mathbf{J}^{-1}\cdot\widetilde{\pmb{\sigma}}^{pt}\cdot(\mathbf{J}^{T}){}^{-1}\cdot\mathbf{S}^{T}\cdot\mathbf{J}^{T}\,,
\end{equation}
where $\widetilde{\pmb{\sigma}}^{pt}$ is the partial transpose of the Williamson form of the original covariance matrix $\pmb{\sigma}$. Let $\mathbf{W}=\mathbf{J}\cdot\mathbf{S}\cdot\mathbf{J}^{-1}$, and we can check that it is not a symplectic matrix, but an orthogonal matrix. Hence $\pmb{\sigma}^{pt}$ is not related to $\widetilde{\pmb{\sigma}}^{pt}$ by a symplectic transformation, and their symplectic eigenvalues do not coincide, not necessarily describing compatible entanglement information. Instead, to find the entanglement measure based on the symplectic eigenvalues, we should find a symplectic matrix $\mathbf{S}'$, which will diagonalize $\pmb{\sigma}^{pt}$ into its own Williamson form $\overline{\pmb{\sigma}}^{pt}$
\begin{align}
	 \overline{\pmb{\sigma}}^{pt}=\mathbf{S}'\cdot\pmb{\sigma}^{pt}\cdot\mathbf{S}'^{T}&=\mathbf{S}'\cdot\mathbf{J}\cdot\pmb{\sigma}\cdot\mathbf{J}^{T}\cdot\mathbf{S}'^{T}\notag\\
	 &=\mathbf{S}'\cdot\mathbf{J}\cdot\mathbf{S}\cdot\widetilde{\pmb{\sigma}}\cdot\mathbf{S}^{T}\cdot\mathbf{J}^{T}\cdot\mathbf{S}'^{T}\\
	 &=\mathbf{S}'\cdot\mathbf{J}\cdot\mathbf{S}\cdot\mathbf{J}^{-1}\cdot\widetilde{\pmb{\sigma}}^{pt}\cdot(\mathbf{J}^{T}){}^{-1}\cdot\mathbf{S}^{T}\cdot\mathbf{J}^{T}\cdot\mathbf{S}'^{T}\neq\widetilde{\pmb{\sigma}}^{pt}\,,\notag
\end{align}
because $\mathbf{S}'\cdot\mathbf{J}\cdot\mathbf{S}\cdot\mathbf{J}^{-1}$ in general is not a symplectic matrix.

Next we turn to the $N$ fully symmetric Gaussian systems; in particular, we stress the connection between the orthogonal matrix in the configuration space and the corresponding symplectic matrix in the phase space. Suppose that the Lagrangian is given by
\begin{equation}
	 \mathcal{L}=\frac{m}{2}\,\mathbf{y}^{T}\cdot\mathbf{y}-\frac{m}{2}\,\mathbf{y}^{T}\cdot\pmb{\Omega}^{2}\cdot\mathbf{y}\,,\label{E:zuwkwz}
\end{equation}
where $\mathbf{y}=(\chi^{(1)},\ldots,\chi^{(N)})^{T}$, and the $N\times N$ interaction matrix is
\begin{equation}
	\pmb{\Omega}^{2}=\begin{pmatrix}\omega_{0}^{2}+\lambda+(N-1)\sigma &-\sigma &\cdots &-\sigma\\
	-\sigma &\omega_{0}^{2}+\lambda+(N-1)\sigma &\cdots &-\sigma\\
	\vdots  &\vdots &\ddots &\vdots\\
	-\sigma &-\sigma &\cdots &\omega_{0}^{2}+\lambda+(N-1)\sigma\end{pmatrix}\,.
\end{equation}
It implies that one of the normal modes must be of the form
\begin{equation}
	\xi^{(1)}=\mathbf{v}^{(1)T}\cdot\mathbf{y}=\frac{1}{\sqrt{N}}\sum_{i=1}^{N}\chi^{(i)}\,,
\end{equation}
corresponding to the eigenfrequency $\eta_{<}^{2}=\omega_{0}^{2}+\lambda$ and the eigenvector
\begin{equation}
	\mathbf{v}^{(1)}=(\frac{1}{\sqrt{N}},\ldots,\frac{1}{\sqrt{N}})^{T}\,.
\end{equation}
The remaining $(N-1)$ normal modes $\xi^{(j)}$ for $j=2,\ldots,N$ are associated with the eigenfrequency of $(N-1)$-fold degeneracy $\eta_{>}^{2}=\omega_{0}^{2}+\lambda+n\sigma>\eta_{<}^{2}$, and $N-1$ orthonormal eigenvectors $\mathbf{v}^{(j)}$. The normal modes $\pmb{\xi}$ are in fact related to the original variables $\mathbf{y}$ by a orthogonal transformation $\mathbf{U}$ such that $\mathbf{y}=\mathbf{U}\cdot\pmb{\xi}$. This $N\times N$ transformation matrix is constructed by the eigenvectors of the interaction matrix $\pmb{\Omega}^{2}$,
\begin{equation}
	\mathbf{U}=(\mathbf{v}^{(1)}\,\mathbf{v}^{(2)}\,\cdots\,\mathbf{v}^{(N)})\,,
\end{equation}
and can be used to diagonalized $\pmb{\Omega}^{2}$,
\begin{equation}
	 \pmb{\Lambda}^{2}=\mathbf{U}^{T}\cdot\pmb{\Omega}^{2}\cdot\mathbf{U}=\operatorname{diag}(\eta_{<}^{2},\eta_{>}^{2},\ldots,\eta_{>}^{2})\,.
\end{equation}
Let the covariance matrix in terms of the normal modes is given by
\begin{equation}
	\widetilde{\pmb{\sigma}}=\begin{pmatrix}\dfrac{1}{2m\eta_{<}}&0\\0 &\dfrac{m\eta_{<}}{2}\end{pmatrix}\oplus\bigoplus_{j=2}^{N}\begin{pmatrix}\dfrac{1}{2m\eta_{>}}&0\\0 &\dfrac{m\eta_{>}}{2}\end{pmatrix}\,.
\end{equation}
We may find the corresponding covariance matrix $\pmb{\sigma}$ in terms of the original variables by an appropriate $2N\times2N$ symplectic transformation matrix $\mathbf{S}$,
\begin{equation}
	\mathbf{S}=\begin{cases}
					S_{2i-1,\,2j-1}=U_{ij}\,,&i,\,j=1,\ldots,N\,,\\
					S_{2i,\,2j}=U_{ij}\,,&i,\,j=1,\ldots,N\,,\\
					0\,,&\text{otherwise}\,,
				\end{cases}
\end{equation}
so that
\begin{align}
	 \mathbf{R}&=\mathbf{S}\cdot\pmb{\mathfrak{R}}\,,&&\text{and}&\pmb{\sigma}&=\mathbf{S}\cdot\widetilde{\pmb{\sigma}}\cdot\mathbf{S}^{T}\,,
\end{align}
if $\mathbf{R}=(\chi^{(1)},p^{(1)},\ldots,\chi^{(N)},p^{(N)})^{T}$ and $\pmb{\mathfrak{R}}=(\xi^{(1)},\pi^{(1)},\ldots,\xi^{(N)},\pi^{(N)})^{T}$ with $p$, $\pi$ being the conjugated momenta to $\chi$, $\xi$, respectively. It is just the compact form for the statements
\begin{align}
	\chi^{(i)}&=U_{ij}\xi^{(j)}\,,&p^{(i)}&=U_{ij}\pi^{(j)}\,,&&\text{for $i,\,j=1,\ldots,N$}\,.
\end{align}
It then implies that
\begin{equation}
	\langle\,\{\chi^{(i)},p^{(j)}\}\,\rangle=U_{ik}U_{jl}\langle\,\{\xi^{(k)},\pi^{(l)}\}\,\rangle=0\,,
\end{equation}
because there is no correlation between the conjugated pair of the normal mode $\xi$; on the other hand, the cross-correlation between $\chi^{(i)}$, $\chi^{(j)}$ is
\begin{equation}
	 \frac{1}{2}\,\langle\,\{\chi^{(i)},\chi^{(j)}\}\,\rangle=U_{ik}U_{jl}\,\frac{1}{2}\,\langle\,\{\xi^{(i)},\xi^{(j)}\}\,\rangle=U_{ik}U_{jl}\,\Delta_{k}\delta_{kj}=U_{ik}U_{jk}\,\Delta_{k}\,,
\end{equation}
where $\Delta_{k}$ is the position uncertainty of the normal mode $\xi^{(k)}$,
\begin{equation}
	\Delta_{k}=\frac{1}{2}\,\langle\,\{\xi^{(k)},\xi^{(k)}\}\,\rangle\,.
\end{equation}
It will take two different values, depending whether $k=1$ or not, so we have
\begin{equation}
	 \frac{1}{2}\,\langle\,\{\chi^{(i)},\chi^{(j)}\}\,\rangle=U_{i1}U_{j1}\,\Delta_{k}+U_{im}U_{jm}\,\Delta_{m}=U_{i1}U_{j1}\,\Delta_{<}+U_{im}U_{jm}\,\Delta_{>}\,,
\end{equation}
where $m=2,\ldots,N$ and
\begin{align}
	\Delta_{<}&=\frac{1}{2m\eta_{<}}\,,&\Delta_{>}&=\frac{1}{2m\eta_{>}}\,,&\Delta_{<}>\Delta_{>}\,.
\end{align}
The orthogonal matrix $\mathbf{U}$ has the property that
\begin{equation}
	U_{i1}U_{j1}+U_{im}U_{jm}=\delta_{ij}\,.
\end{equation}
Therefore we arrive at
\begin{equation}
	 \frac{1}{2}\,\langle\,\{\chi^{(i)},\chi^{(j)}\}\,\rangle=\delta_{ij}\,\Delta_{>}-U_{i1}U_{j1}\bigl(\Delta_{>}-\Delta_{<}\bigr)=\delta_{ij}\,\Delta_{>}-\frac{1}{N}\,\bigl(\Delta_{>}-\Delta_{<}\bigr)\,,
\end{equation}
since $U_{i1}=(\mathbf{v}^{(1)})_{i}=1/\sqrt{N}$. Similarly we have
\begin{equation}
	 \frac{1}{2}\,\langle\,\{p^{(i)},p^{(j)}\}\,\rangle=\delta_{ij}\,\Theta_{>}-\frac{1}{N}\,\bigl(\Theta_{>}-\Theta_{<}\bigr)\,,
\end{equation}
with $\Theta_{k}=\dfrac{1}{2}\,\langle\,\{\pi^{(k)},\pi^{(k)}\}\,\rangle$,
\begin{align}
	\Theta_{<}&=\frac{m\eta_{<}}{2}\,,&\Theta_{>}&=\frac{m\eta_{>}}{2}\,,&\Theta_{>}>\Theta_{<}\,.
\end{align}
This tells us that generically the covariance matrix $\pmb{\sigma}$ takes the form
\begin{align}\label{E:dfnekr}
	\pmb{\sigma}&=\begin{pmatrix}\pmb{\beta} &\pmb{\epsilon} &\cdots &\pmb{\epsilon}\\\pmb{\epsilon} &\pmb{\beta} &\cdots &\pmb{\epsilon}\\\vdots &\vdots &\ddots &\vdots\\\pmb{\epsilon} &\pmb{\epsilon} &\cdots &\pmb{\beta}\end{pmatrix}\,,&\pmb{\beta}&=\begin{pmatrix}a &0 \\[4pt]0 &e\end{pmatrix}\,,&\pmb{\epsilon}&=\begin{pmatrix}b &0 \\[4pt]0 &f\end{pmatrix}\,,
\end{align}
and now
\begin{align}
	 a&=\Delta_{>}-\frac{1}{N}\,\bigl(\Delta_{>}-\Delta_{<}\bigr)>0\,,&e&=\Theta_{>}-\frac{1}{N}\,\bigl(\Theta_{>}-\Theta_{<}\bigr)>0\,,\\
	b&=-\frac{1}{N}\,\bigl(\Delta_{>}-\Delta_{<}\bigr)>0\,,&f&=-\frac{1}{N}\,\bigl(\Theta_{>}-\Theta_{<}\bigr)<0\,.
\end{align}
We convert it to the standard form
\begin{equation}
	\pmb{\sigma}'=\mathbf{V}\cdot\pmb{\sigma}\cdot\mathbf{V}^{T}=\begin{pmatrix}\pmb{\beta}' &\pmb{\epsilon}' &\cdots &\pmb{\epsilon}'\\\pmb{\epsilon}' &\pmb{\beta}' &\cdots &\pmb{\epsilon}'\\\vdots &\vdots &\ddots &\vdots\\\pmb{\epsilon}' &\pmb{\epsilon}' &\cdots &\pmb{\beta}'\end{pmatrix}\,,
\end{equation}
with
\begin{align}
	\pmb{\beta}'&=\mathbf{K}\cdot\pmb{\beta}\cdot\mathbf{K}^{T}=\begin{pmatrix}\lambda &0 \\[4pt]0 &\lambda\end{pmatrix}\,,&\pmb{\epsilon}'&=\mathbf{K}\cdot\pmb{\epsilon}\cdot\mathbf{K}^{T}=\begin{pmatrix}\zeta_{1} &0 \\[4pt]0 &\zeta_{2}\end{pmatrix}\,,
\end{align}
by the symplectic matrix $\mathbf{V}$
\begin{align}
	 \mathbf{V}&=\bigoplus_{k=1}^{N}\mathbf{K}\,,&\mathbf{K}&=\operatorname{diag}(\sqrt[4]{\frac{e}{a}},\sqrt[4]{\frac{a}{e}})\,.
\end{align}
Therefore we have
\begin{align}
	\lambda&=\sqrt{ae}\,,&\zeta_{1}&=b\sqrt{\frac{e}{a}}\,,&\zeta_{2}&=f\sqrt{\frac{a}{e}}\,.
\end{align}
Since it has been shown in~\cite{SAI05} that for such a fully symmetric system, there are only two distinct symplectic eigenvalues for the covariance matrix $\pmb{\sigma}'$. They are
\begin{equation*}
	\eta_{+}=\sqrt{\bigl[\lambda+\bigl(N-1\bigr)\zeta_{1}\bigr]\bigl[\lambda+\bigl(N-1\bigr)\zeta_{2}\bigr]}\,,
\end{equation*}
and the $(N-1)$-fold degenerate
\begin{equation*}
	\eta_{-}=\sqrt{\bigl(\lambda-\zeta_{1}\bigr)\bigl(\lambda-\zeta_{2}\bigr)}\,.
\end{equation*}
From we have derived so far, we obtain
\begin{align}
	\lambda+\bigl(N-1\bigr)\zeta_{1}&=\sqrt{\frac{e}{a}}\Bigl[a+(N-1)b\Bigr]=\sqrt{\frac{e}{a}}\,\Delta_{<}\,,\\
	\lambda+\bigl(N-1\bigr)\zeta_{2}&=\sqrt{\frac{a}{e}}\Bigl[e+(N-1)f\Bigr]=\sqrt{\frac{a}{e}}\,\Theta_{<}\,,\\
	\lambda-\zeta_{1}&=\sqrt{\frac{e}{a}}\Bigl[a-b\Bigr]=\sqrt{\frac{e}{a}}\,\Delta_{>}\,,\\
	\lambda-\zeta_{2}&=\sqrt{\frac{a}{e}}\Bigl[e-f\Bigr]=\sqrt{\frac{a}{e}}\,\Theta_{>}\,.
\end{align}
Hence these two symplectic eigenvalues take rather simple form
\begin{align}
	\eta_{+}&=\sqrt{\Delta_{<}\Theta_{<}}\,,&\eta_{-}&=\sqrt{\Delta_{>}\Theta_{>}}\,,
\end{align}
and both of them are exactly $1/2$. This is the consequence of the uncertainty principle for the ground state. If the states for the normal modes are not their ground states, then both $\eta_{-}$, $\eta_{+}$ should be greater than $1/2$.

\end{document}